\documentclass{article}

\usepackage[preprint]{neurips_2024}
\usepackage{fancyhdr} 
\usepackage{blindtext} 
\usepackage{makecell}
\usepackage{natbib}

\usepackage{fontawesome5} 

\usepackage[utf8]{inputenc} 
\usepackage[T1]{fontenc}    
\usepackage{hyperref}       
\usepackage{url}            
\usepackage{booktabs}       
\usepackage{amsfonts}       
\usepackage{nicefrac}       

\usepackage{xcolor}         
\usepackage{amsmath}
\usepackage{float}
\usepackage{tabularray}
\usepackage{multirow}
\usepackage{booktabs, caption, array}
\usepackage{amssymb}
\usepackage{lipsum}         
\usepackage{CJKutf8}        
\usepackage{nomencl}
\usepackage{glossaries}
\usepackage{xspace}
\usepackage{multirow}
\usepackage{graphicx}
\usepackage{bbding}
\usepackage{pifont}
\usepackage{algorithm}
\usepackage{enumitem}
\usepackage{multicol}
\usepackage{caption}
\usepackage{array}
\usepackage{fp}
\usepackage{multirow}   
\usepackage{makecell}
\usepackage{flushend}
\usepackage{cases}
\usepackage{color}
\usepackage{algpseudocode}
\usepackage{listings}
\usepackage{mathrsfs}
\usepackage{longtable}
\usepackage{colortbl}
\usepackage{bm}
\usepackage{caption}
\usepackage{natbib}
\usepackage{tabularx}

\usepackage{subcaption} 

\setlist[itemize]{noitemsep, topsep=0pt}
\usepackage{arydshln}
\usepackage{lipsum}
\definecolor{codegreen}{rgb}{0,0.3,0.6}
\definecolor{codegray}{rgb}{0.5,0.5,0.5}

\newcommand{\ie}{\emph{i.e.,}\xspace}
\newcommand{\eg}{\emph{e.g.,}\xspace}

\newcommand{\paratitle}[1]{\vspace{1.5ex}\noindent\textbf{#1}}

\newcommand{\ignore}[1]{}

\usepackage[most]{tcolorbox} 
\usepackage{microtype}      
\definecolor{darkorange}{RGB}{255, 140, 0}
\definecolor{lightgreen}{RGB}{145, 204, 117}
\definecolor{lightyellow}{RGB}{250, 200, 88}
\definecolor{lightred}{RGB}{238, 102, 102}
\definecolor{lightblue}{RGB}{115, 192, 222}
\newtcolorbox{promptbox}[3][Judge Prompt]{
colback=black!5!white,
arc=5pt, 
boxrule=0.5pt,
fonttitle=\bfseries,
title=#1, 
before upper={\small}, fontupper=\fontfamily{ptm}\selectfont,
colframe=#2,
label=#3,
}
\definecolor{gray_1}{HTML}{B7B7B7}
\definecolor{gray_2}{HTML}{F0F0F0} 
\definecolor{frame_blue}{HTML}{A9D18E}

\newtcolorbox[auto counter, number within=section]{PromptBoxNew}[2][]{
    enhanced,
    breakable,
    colback=gray_2, 
    colframe=gray_1,
    coltitle=white,
    fontupper=\small,
    fonttitle=\bfseries,
    title={#2}, 
    label={#1},
    arc=2pt,
    boxrule=1pt,
    left=2mm, right=2mm, top=2mm, bottom=2mm,
}

\usepackage{etoolbox}
\makeatletter
\patchcmd{\@maketitle}
  {\@toptitlebar}
  {%
    \hbox to \textwidth{%
    
    \includegraphics[height=0.76cm]{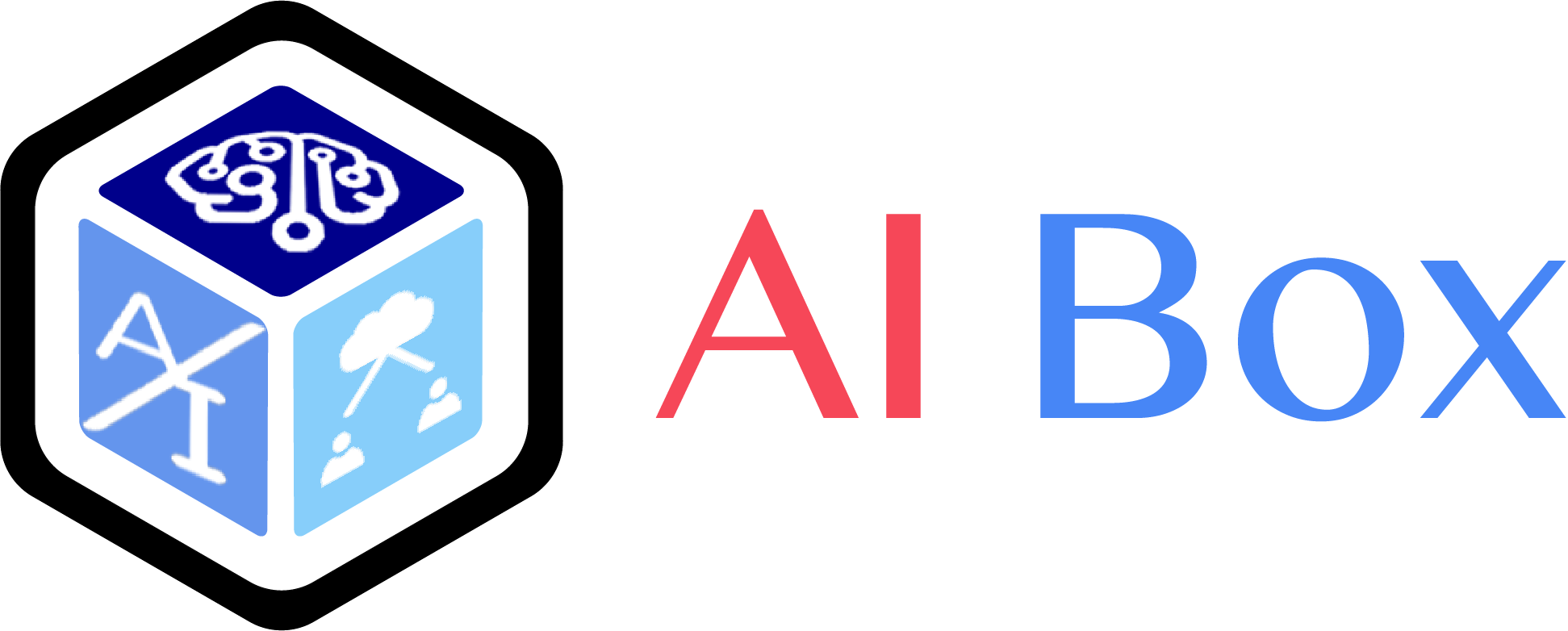}%
      \hfill
      \includegraphics[height=0.7cm]{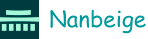}%
    }%
    \vspace{0.3em}
    \@toptitlebar
  }
  {}{}
\makeatother

\title{SWE-Master: Unleashing the Potential of Software Engineering Agents via  Post-Training}

\author{%
   Huatong Song$^{1}$\thanks{Equal contribution.}~, 
  Lisheng Huang$^{1*}$, 
  Shuang Sun$^{1*}$, 
  Jinhao Jiang$^{1*}$,
  Ran Le$^2$,\\
  \textbf{Daixuan Cheng$^{1}$},
  \textbf{Guoxin Chen$^{1}$},
  \textbf{Yiwen Hu$^{1}$, Zongchao Chen$^{2}$, Yiming Jia$^{2}$,}\\
  \textbf{Wayne Xin Zhao$^{1}$\thanks{Correspondence to Wayne Xin Zhao and Yang Song.}~,} 
  \textbf{ Yang Song$^{2\dagger}$, } 
   \textbf{{Tao Zhang}$^{2}$, Ji-Rong Wen$^{1}$}
  \\
  $^1$Gaoling School of Artificial Intelligence, Renmin University of China.\\
  $^2$BOSS Zhipin, Beijing, China.\\
  \texttt{songhuatong123@ruc.edu.cn},
  \texttt{batmanfly@gmail.com}, \texttt{songyang@kanzhun.com}
}

\begin{document}
\maketitle




\definecolor{babyblue}{HTML}{F8F9FE}
\newtcolorbox{bluebox}{
  colback=babyblue,    
  colframe=babyblue,  
  width=1.0\textwidth,  
  center,               
  arc=8pt,                 
  boxrule=0pt,           
  boxsep=0pt,       
  left=2pt,               
  right=2pt,              
  top=10pt,                
  bottom=10pt              
}


\begin{bluebox}
\begin{abstract}

In this technical report, we present SWE-Master, an open-source and fully reproducible post-training framework for building effective software engineering agents. SWE-Master systematically explores the complete agent development pipeline, including teacher-trajectory synthesis and data curation, long-horizon SFT, RL with real execution feedback, and inference framework design. Starting from an open-source base model with limited initial SWE capability, SWE-Master demonstrates how systematical optimization method can elicit strong long-horizon SWE task solving abilities. We evaluate SWE-Master on SWE-bench Verified, a standard benchmark for realistic software engineering tasks. Under identical experimental settings, our approach achieves a resolve rate of 61.4\% with Qwen2.5-Coder-32B, substantially outperforming existing open-source baselines. By further incorporating test-time scaling~(TTS) with LLM-based environment feedback, SWE-Master reaches 70.8\% at TTS@8, demonstrating a strong performance potential. SWE-Master provides a practical and transparent foundation for advancing reproducible research on software engineering agents. The code is available at \url{https://github.com/RUCAIBox/SWE-Master}.

\end{abstract}
\end{bluebox}

\begin{figure}[H]
\centering
    \centering
    \includegraphics[width=0.98\linewidth]{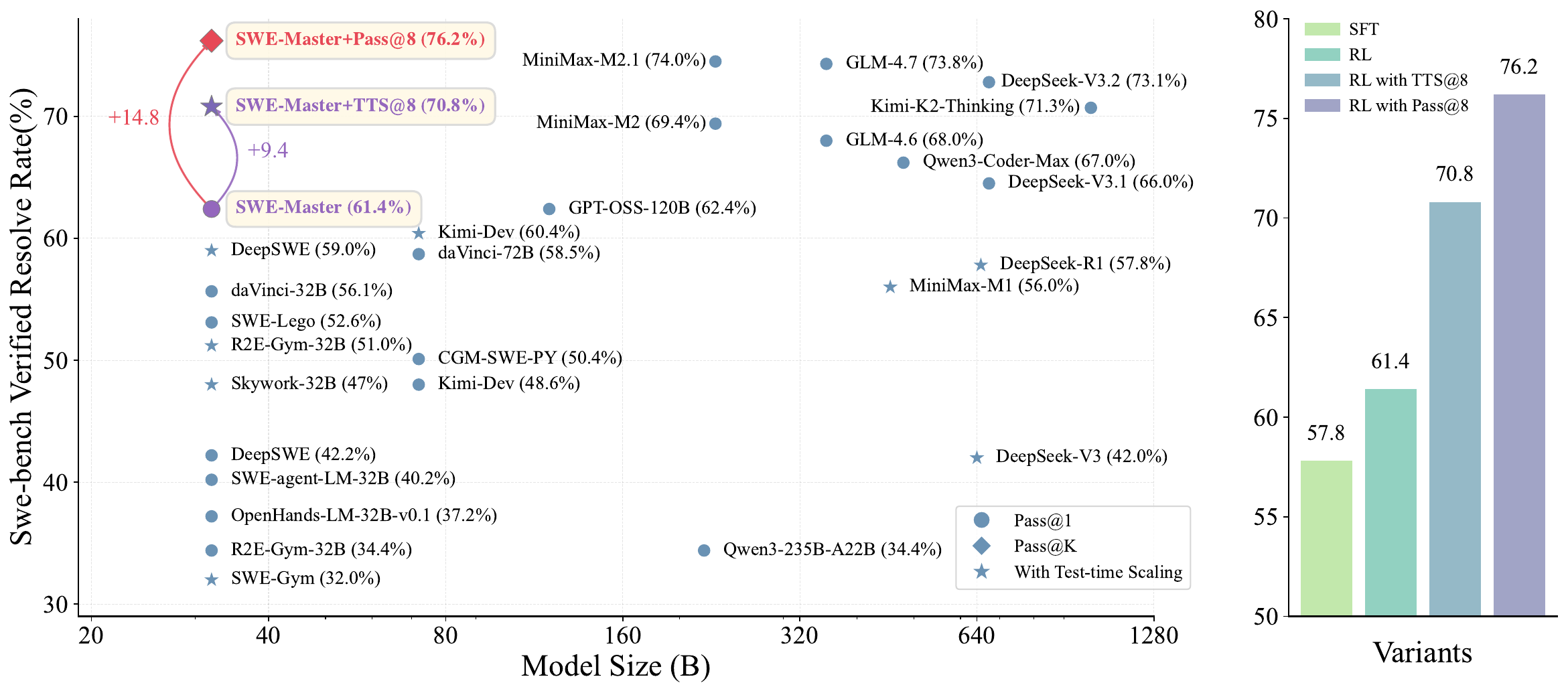}
\caption{ Performance overview and scaling analysis of SWE-Master. \textbf{Left:} Comparasion of the perference of various open-source foundational models and SWE agents on SWE-bench Verified. \textbf{Right:} Performance of SWE-Master across different training stages and evaluation metrics. }
    \label{fig:benchmarks}
\end{figure}

\section{Introduction}
\label{sec-intro}
Large language model based software engineering agents, also referred to as \textit{SWE agents}~\citep{li2026advances}, have recently emerged as a powerful paradigm for automating complex software development tasks~\citep{yang2024swe, jimenez2024swebench}. Unlike traditional code generation models that focus on short snippets or isolated functions~\citep{hui2024qwen2, guo2024deepseek}, modern SWE agents are expected to understand natural language requirements, navigate large codebases, modify multiple files, execute tests, and iteratively refine solutions until a task is successfully completed~\citep{wang2025openhands}. By operating at the level of end-to-end autonomous workflows, SWE agents have the potential to significantly reduce human engineering effort and accelerate software development and maintenance in real-world settings.

Recent progress in SWE agents has been driven by coordinated advances across the entire pipeline, spanning data construction, training with environment feedback, and inference-time scaffold. On the training side, mainstream approaches construct executable task instances from real-world GitHub issues and train models as agents that interact with environments over multiple steps—exploring codebases, modifying files, executing commands, and iteratively refining solutions until a final patch is validated by unit tests, with execution feedback obtained from containerized execution environments~(\ie Docker) providing supervision signals~\citep{sonwane2025bugpilot, cao2025skyrl, cheng2026llm, golubev2025training}. On the inference side, existing methods typically adopt standardized scaffolds with basic capability workflows, such as OpenHands~\citep{wang2025openhands}. Some studies further augment these frameworks with additional tools to support extended capabilities, including long-context management~\citep{liu2025context, wang2026swe, sun2025scaling}. Through systematic training optimization combined with well-designed inference frameworks, recent systems developed by organizations such as OpenAI and Anthropic have achieved strong performance on challenging real-world software engineering benchmarks~\citep{openai_gpt51_codex_max, anthropic_claude_sonnet_4_5}.

Despite the rapid progress of software engineering agents, existing approaches remain fundamentally limited by the lack of transparency and reproducibility across training data construction and optimization procedures. In practice, the closed nature of many state-of-the-art systems obscures several critical challenges that are essential for building effective SWE agents. On the training data side, a key difficulty lies in efficiently constructing high-quality teacher trajectories that capture long-horizon reasoning and realistic environment interactions. On the optimization side, agent training typically follows a two-stage paradigm: SFT and RL. The former requires careful data filtering and mixture design to balance correctness, diversity, and task difficulty, while the latter demands delicate algorithms tuning and reward formulations to encourage sufficient exploration and stable learning, without suffering from issues such as entropy collapse or reward hacking. In addition, on the inference side, existing approaches are largely constrained by basic agent frameworks, with limited exploration of advanced tools and system designs, particularly in terms of execution efficiency and long-context management. Together, these opaque and interdependent components form a high barrier to entry, hindering reproducible research and limiting the accessibility of SWE agent development for the broader academic community.

To address these challenges, we introduce SWE-Master, an open-source software engineering agent framework that fully exposes the post-training pipeline in a transparent and reproducible manner. Rather than treating agent performance as the outcome of isolated design choices, SWE-Master systematically studies how software engineering capabilities emerge from the interaction between data construction, optimization strategies, and inference-time behaviors, even when starting from an open-source model with limited initial SWE task performance (\eg below 10 points on SWE-bench Verified benchmarks using Qwen2.5-Coder-32B model)~\citep{hui2024qwen2}. In particular, we analyze the impact of different teacher models and data filtering strategies during trajectory synthesis, and show that controlling the difficulty distribution of training data plays a crucial role in shaping the interaction depth and decision-making behavior of models after SFT. Building on this foundation, we further investigate RL in real execution environments by exploring combinations of optimization algorithms and reward designs, enabling efficient exploration and effective learning while mitigating common failure modes such as reward hacking and unstable behaviors. Together, SWE-Master provides a comprehensive, open, and empirically grounded framework for understanding and advancing the post-training of software engineering agents.

Building on the previously discussed limitations of existing inference frameworks, we further investigate the impact of equipping advanced capabilities at inference time. Motivated by the observation that many software engineering failures stem from insufficient understanding of large codebases rather than code generation errors, we focus on enhancing agents’ code interaction and navigation abilities. In particular, we study the transition from simple text-based search to structured code navigation based on language server protocols, and analyze its impact on reasoning and decision making in large repositories. 
Through systematic empirical analysis, we find that tools grounded in the Language Server Protocol (LSP) constitute a new foundational paradigm for SWE agents. This approach empowers agents with IDE-grade code comprehension, thereby facilitating precise inspection and modification of complex file systems within realistic software engineering scenarios.

To validate the effectiveness of the proposed approach, we conduct extensive experiments on SWE-bench Verified~\citep{sweb-verified}, a widely used benchmark for evaluating realistic software engineering agents. Under identical experimental settings, including the same base model, training data sources, and inference configurations, our long-horizon SFT strategy significantly outperforms existing open-source methods, achieving a resolve rate of 57.8\%. These results indicate that careful data curation and trajectory-level supervision alone can substantially improve performance on real-world software engineering tasks. Building on this strong SFT baseline, we further apply RL with real execution environments, which consistently extends model capabilities and enables the agent to solve more challenging instances, pushing the performance to 61.4\%.
Furthermore, inspired by prior studies that leverage LLMs to simulate real execution feedback~\cite{shum2025swe, jain2025r2e}, we adopt a test-time scaling (TTS) strategy~\citep{swe-world} powered by LLM-based environment feedback. This approach enables the agent to explore and rank multiple candidate solutions without incurring the overhead of physical execution. By selecting the most promising candidate, our method achieves a score of 70.8\% under the TTS@8 setting.
This strategy avoids direct execution in real environments, which is particularly valuable in scenarios where environment interactions are costly, irreversible, or unsafe. Finally, by integrating an LSP-based code navigation framework at inference time, SWE-Master improves agent efficiency with minimal impact on task success rates, achieving a practical balance between effectiveness and efficiency.

Based on our experiments, our major contritions are summarized below:

$\bullet$ We release the first fully open-source, end-to-end training pipeline for software engineering agents, covering data processing, SFT, RL infrastructure and strategies, and inference-time agent frameworks.

$\bullet$ We introduce IDE-level capabilities based on LSP–driven code navigation, enabling more efficient and structured repository understanding, and significantly improving agent efficiency without sacrificing performance.

$\bullet$ We significantly advance open-source model performance on SWE-bench Verified, achieving 61.4\% accuracy with Qwen2.5-Coder-32B, improving to 70.8\% with test-time scaling and 76.2\% under Pass@8, demonstrating strong perpormance potential.
\section{Preliminaries}
\label{sec:preliminaries}


\subsection{Problem Formulation: The SWE Task}
We define the software engineering task as an automated program repair or feature implementation problem. Formally, let $\mathcal{D} = \{(I_i, \mathcal{C}_i, \mathcal{U}_i)\}_{i=1}^N$ represent a dataset of software engineering problems. For a specific instance, the input consists of:
\begin{itemize}[leftmargin=2.0em]
    \item An issue description $I$, which describes the bug report or the feature request.
    \item A codebase $\mathcal{C}$, representing the initial state of the code repository (\ie the file system structure).
\end{itemize}
The ground truth typically includes a golden patch $p^*$ and a unit test suite $\mathcal{U}$ comprising a series of test cases. The goal of the model is to generate a patch $\hat{p}$ (a set of diffs) such that applying the patch to the codebase resolves the issue $I$, defined as effectively passing all unit tests. Let $f_{\text{apply}}(\mathcal{C}, p)$ be a function that applies a patch to the codebase. The modified codebase is denoted as $\mathcal{C}' = f_{\text{apply}}(\mathcal{C}, \hat{p})$.

\subsection{Agent-Based Environment Interaction}
We formulate the problem solving process as a sequential decision-making process within an interactive environment. The environment state at step $k$ is denoted as $s_k$, which includes the current file contents, the command line history, and the previous execution outputs.

The agent functions as a policy $\pi_\theta(a_k | h_k)$, where $\theta$ represents the model parameters and $h_k = \langle s_0, a_0, o_0, s_1, \dots, o_{k-1} \rangle$ is the interaction history. The agent generates an action $a_k$ consisting of a reasoning trace (Thought) and a tool invocation (Action). The action space $\mathcal{A}$ typically includes:
\begin{equation}
    \mathcal{A} = \mathcal{A}_{\text{nav}} \cup \mathcal{A}_{\text{edit}} \cup \mathcal{A}_{\text{exec}}
\end{equation}
where $\mathcal{A}_{\text{nav}}$ contains navigation commands (\eg \texttt{ls}, \texttt{cd}), $\mathcal{A}_{\text{edit}}$ contains file manipulation commands (\eg  \texttt{view}, \texttt{create}, \texttt{str\_replace}), and $\mathcal{A}_{\text{exec}}$ contains execution commands (\eg \texttt{pytest}).

Upon executing action $a_k$, the environment returns an observation $o_{k}$ (\eg standard output, error logs, or file content) and transitions to a new state $s_{k+1}$. The agent then proceeds with subsequent interactions.This process continues until the agent issues a termination action (\eg \texttt{submit}) or reaches a maximum step limit $K_{\text{max}}$. The trajectory is defined as $\tau = \langle I, a_0, o_0, a_1, o_1, \dots, a_K \rangle$.

\subsection{Evaluation Protocol}\label{sec:preliminaries_verify}
Evaluation is strictly execution-based within isolated Docker containers. 
For each issue, the model interacts with the codebase to implement a solution, which is subsequently captured as a patch via \texttt{git diff}. This patch is then applied to the original repository for verification.
The validity of a generated patch $\hat{p}$ is determined by the unit test suite $\mathcal{U}$. The test suite consists of two subsets: $\mathcal{U} = \mathcal{U}_{\text{fail}} \cup \mathcal{U}_{\text{pass}}$. 
Specifically, $\mathcal{U}_{\text{fail}}$ denotes the set of \textit{fail-to-pass} (F2P) tests designed to reproduce the bug, whereas $\mathcal{U}_{\text{pass}}$ comprises \textit{pass-to-pass} (P2P) tests intended to ensure no regression in existing functionality.

Let $V(\mathcal{C}, u_j)$ be the verification reward function for a unit test $u_j \in \mathcal{U}$, defined such that $V(\mathcal{C}, u_j) = 1$ if $u_j$ passes on codebase $\mathcal{C}$, and $0$ otherwise. A software engineering task is considered resolved if and only if the modified codebase $\mathcal{C}'$ successfully passes the entire test suite $\mathcal{U}$, where $\mathcal{C}'$ is obtained by applying the predicted patch $\hat{p}$ to the original codebase $\mathcal{C}$. Formally, the resolution status $\text{Resolved}(\hat{p})$ is defined as:
\begin{equation}
    \text{Resolved}(\hat{p}) = \mathbb{I} \left[ \sum_{u_j \in \mathcal{U}} V(\mathcal{C}', u_j) = |\mathcal{U}| \right]
\end{equation}
where $\mathbb{I}[\cdot]$ denotes the indicator function. Consequently, the task-level reward is unity if and only if the verification reward is 1 for every individual test case $u_j \in \mathcal{U}$; otherwise, the reward remains 0.


\section{SWE-Master: Training Open-Source SWE Agent}
\label{method}

\subsection{Training Framework and Environments}\label{sec:framework_and_env}
Training effective issue-solving code agents requires environments that closely reflect real-world Software Engineering workflows. Unlike static benchmarks (\eg code genreation~\citep{jain2024livecodebench}, websearch~\citep{wei2025browsecomp}), such tasks demand interactive execution environments with terminal access, persistent file systems, and package management support, allowing agents to compile, run, and debug code under realistic conditions. To enable reliable trajectory collection and maintain stability for SFT and RL, we apply a robust and systematic framework for environment interaction.

The overall inference pipeline is based on R2E-Gym framework~\citep{jain2025r2e}, which is a lightweight scaffold adapted from OpenHands and follows a standard ReAct-style interaction loop~\citep{react}. To support this interaction logic, we adopt a decoupled Docker–Server architecture, where execution environments are deployed on dedicated CPU nodes, thereby remaining physically separated from model inference servers. This design enables the on-demand creation of lightweight and isolated coding environments while ensuring stable and uninterrupted inference. Each container provides the essential components required for agent training, including a terminal interface, a file system, and the corresponding code repositories. Network access is preserved to support standard package installation and dependency management.
Our framework integrates several widely used open-source SWE Python datasets that rely on Docker, including SWE-Gym~\citep{swe-gym}, R2E-Gym~\citep{jain2025r2e}, SWE-smith~\citep{yang2025swesmith}, and SWE-rebench~\citep{badertdinov2025swerebench}. All unit tests are built offline and preloaded into their respective Docker images before evaluation. Given the large number of Docker images involved (approximately 13,000), we distribute them across multiple CPU nodes. During inference, requests are routed according to the associated issue identifier to locate the appropriate node and initialize the required environment.

For each issue, the agent interacts with the environment through a set of tools: \texttt{bash\_execute}, \texttt{file\_editor}, and \texttt{submit}. These tools provide the functionality necessary for resolving software issues. In addition, we support higher-level tools built on the Language Server Protocol (LSP)~\citep{lsp}, which are described in Section~\ref{sec:ide_injection}.
To preserve evaluation integrity and mitigate the risk of \textit{git hacking}~\citep{xiao2026mimo}, we enforce strict security constraints within the execution environments. In particular, the potentially exploitable \texttt{git}-related commands (\ie \texttt{git log} and \texttt{git show}) are disabled to prevent the agent from accessing remote repositories or retrieving ground-truth solutions, thereby reducing the risk of data leakage.

By combining physical isolation of Dockers with a decoupled server design, the system sustains efficient policy inference under high concurrency. This infrastructure supports large-scale parallel data collection and stable RL training, making it well suited for scalable code agent development.

\subsection{Trajectory Synthesis and Data Curation}

This section outlines the trajectory generation and fine-grained filtering pipeline for the SWE dataset. Table~\ref{tab:sft_data_dist} summarizes the statistics of candidate datasets and rollouts, while Figure~\ref{fig:data_filter_dist} illustrates the data distribution following the applied filtration strategies.

\subsubsection{Agent-Based Trajectory Rollout}\label{sec:traj_gen}

As described in Section~\ref{sec:framework_and_env}, we adopt multiple established software engineering datasets that are packaged with Docker environments, including SWE-Gym, SWE-rebench, R2E-Gym, and SWE-smith. We use MiniMax-M2~\citep{minimax_m2} and GLM-4.6~\citep{glm_46} as teacher models to  generate trajectories, using the inference pipeline based on R2E-Gym framework. The rollout process follows an agent-based paradigm, in which the model interacts directly with a realistic execution environment. Specifically, the agent is able to explore the target code repository, modify source files, write and execute unit tests to validate proposed fixes, and iteratively revise previous changes based on test outcomes.
Throughout this process, we record both the model’s internal reasoning traces and its function call sequences, yielding complete interaction trajectories paired with corresponding reward signals. 
To assess the difficulty of individual issues, we conduct $N$ rollouts (with $N \in [3, 12]$) for each issue and generate $N$ trajectories, that form the foundation for subsequent data filtering and training stages.

Table~\ref{tab:sft_data_dist} summarizes the composition and generation metrics of our rollout data corpus, which integrates a hybrid of real-world and synthetic data sources. The collection demonstrates high structural diversity; notably, SWE-rebench provides extensive repository coverage with over 1,400 unique repos, while SWE-smith contributes a significant volume of samples.

\begin{figure}[ht]
\centering
    \centering
    \includegraphics[width=0.9\linewidth]{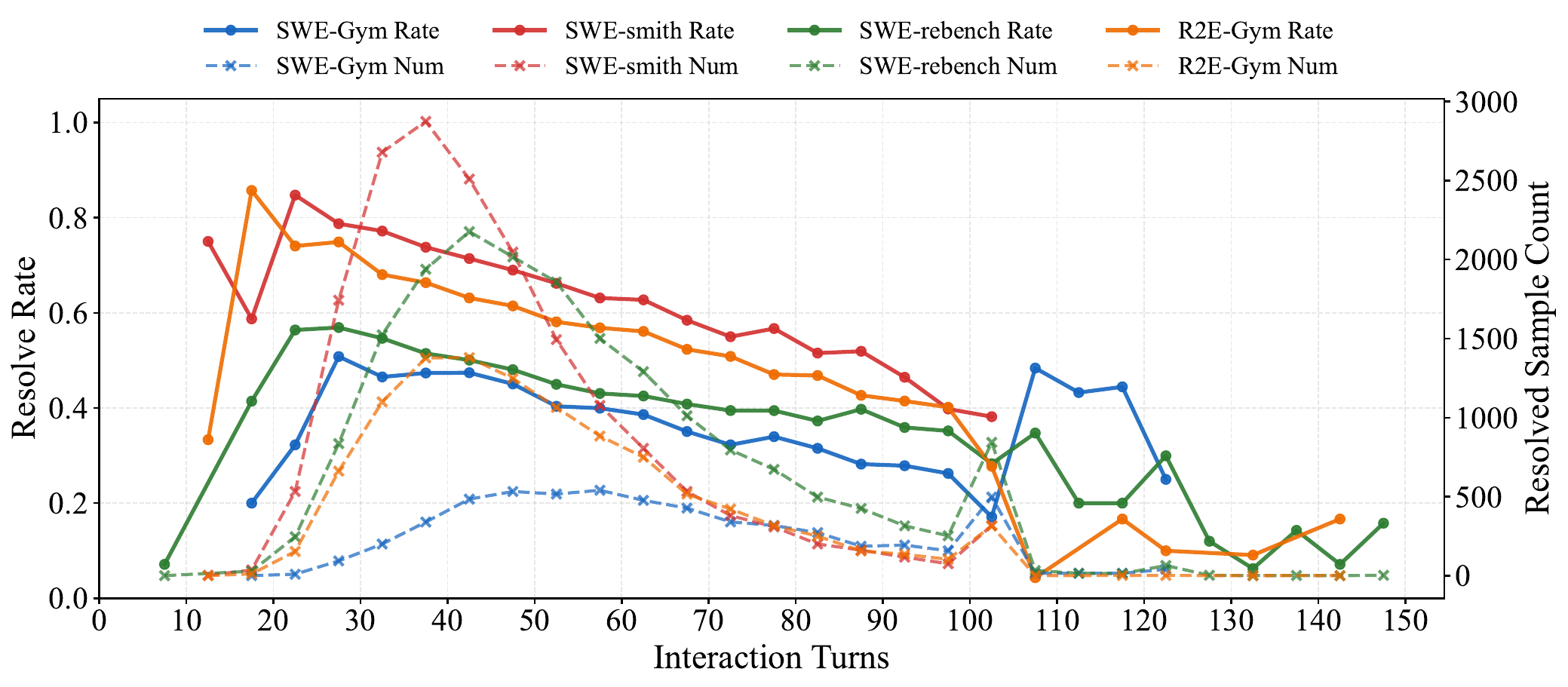}
\caption{Distribution of resolve rates and resolved sample counts across interaction turns for SWE-Gym, SWE-smith, SWE-rebench, and R2E-Gym.}
    \label{fig:four_dataset_rollout}
\end{figure}

Meanwhile, 
Figure~\ref{fig:four_dataset_rollout} illustrates the correlation between interaction turns and model performance across four datasets. A consistent inverse correlation is observed: resolve rates progressively decline as the number of turns increases, indicating that extended interaction budgets fail to guarantee success in more complex issue-solving problem. This persistence of failure stems from both the intrinsic difficulty of the tasks and the accumulation of noise inherent in long-horizon interactions. The distribution of solved samples is predominantly concentrated between 20 and 60 interaction turns, suggesting that the majority of instances are relatively simple and can be successfully resolved with a limited number of interactions.

\begin{table*}[t]
    \small
    \centering
    \caption{Statistics and distribution of open-source SWE data. The right section details the yield from the rollout process and the final number of trajectories selected after filtering for use in SFT.}
    \label{tab:sft_data_dist}
    \setlength{\tabcolsep}{3.2pt} 
    \renewcommand{\arraystretch}{1.2} 

    \begin{tabular}{c c ccc ccc}
        \toprule
        \multirow{2}{*}{\textbf{Dataset}} & \multirow{2}{*}{\textbf{Source}} & \multicolumn{3}{c}{\textbf{Dataset Statistics}} & \multicolumn{3}{c}{\textbf{Generation \& Filtering}} \\
        \cmidrule(lr){3-5} \cmidrule(lr){6-8}
         & & \textbf{\# Samples} & \textbf{\# Images} & \textbf{\# Repos} & \textbf{Res. Inst.} & \textbf{Res. Trajs.} & \textbf{Final Trajs.} \\
        \midrule
        
        {SWE-Gym} & Real & 2,438 & 2,438 & 11 & 1,068 & 5,685 & 2,948 \\
        
        {SWE-rebench} & Real & 6,542 & 6,542 & 1,429 & 4,268 & 10,861 & 7,157 \\
        {R2E-Gym} & Synthetic & 4,578 & 4,578 & 10 & 3,234 & 18,398 & 2,462 \\
        {SWE-smith} & Synthetic & 14,103 & 114 & 114 & 6,353 & 17,901 & - \\
        
        \bottomrule
    \end{tabular}
\end{table*}

\subsubsection{Data Filtering}\label{sec:data_filting}

\paratitle{Format-Based Filter.} We apply a rigorous quality filtration protocol to the raw generated trajectories. First, we eliminate unsuccessful attempts by discarding trajectories with a reward of zero. Second, to ensure computational stability, we prune instances exceeding a context length of 80K tokens or 100 turns as we find that these outliers constitute merely 5\% of the resolved instances yet pose a disproportionate risk of out-of-memory (OOM) errors. Finally, we filter out trajectories containing syntactically invalid actions, specifically defining these as unparsable function calls or erroneous multiple invocations.

\begin{figure}[ht]
\centering
    \centering
    \includegraphics[width=0.90\linewidth]{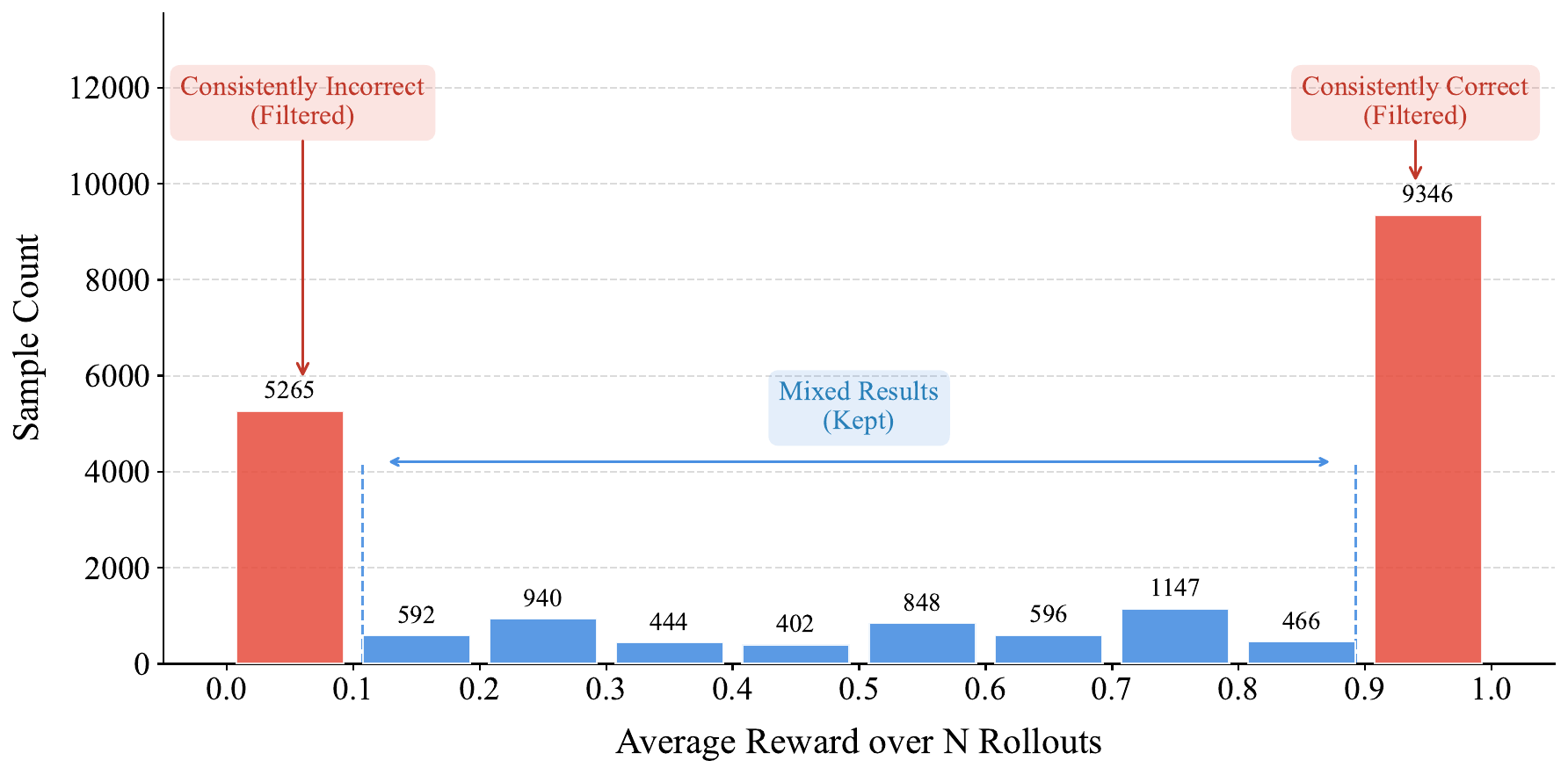}
\caption{Difficulty distribution across SWE datasets estimated via best-of-n performance. The datasets include SWE-Gym, SWE-smith, SWE-rebench, and R2E-Gym.}
    \label{fig:data_rollout_acc_distribution}
\end{figure}

\paratitle{Difficulty-Based Filter.} Existing open-source SWE datasets lack explicit annotations for issue difficulty. As described in Section~\ref{sec:traj_gen}, we conduct $N$ rollouts (with $N \in [3, 12]$) for each issue and compute the average resolve rate, which serves as a proxy for issue difficulty. As illustrated in Figure~\ref{fig:data_rollout_acc_distribution}, the distribution exhibits a bimodal pattern where the majority of issues are either consistently solved (trivial) or consistently failed (intractable). Consequently, we exclude these polar extremes from the candidate pool, selectively retaining only those issues that yield a mixture of successful and failed trajectories to ensure the training set focuses on samples with learnable difficulty.

\begin{figure}[ht]
\centering
    \centering
    \includegraphics[width=0.9\linewidth]{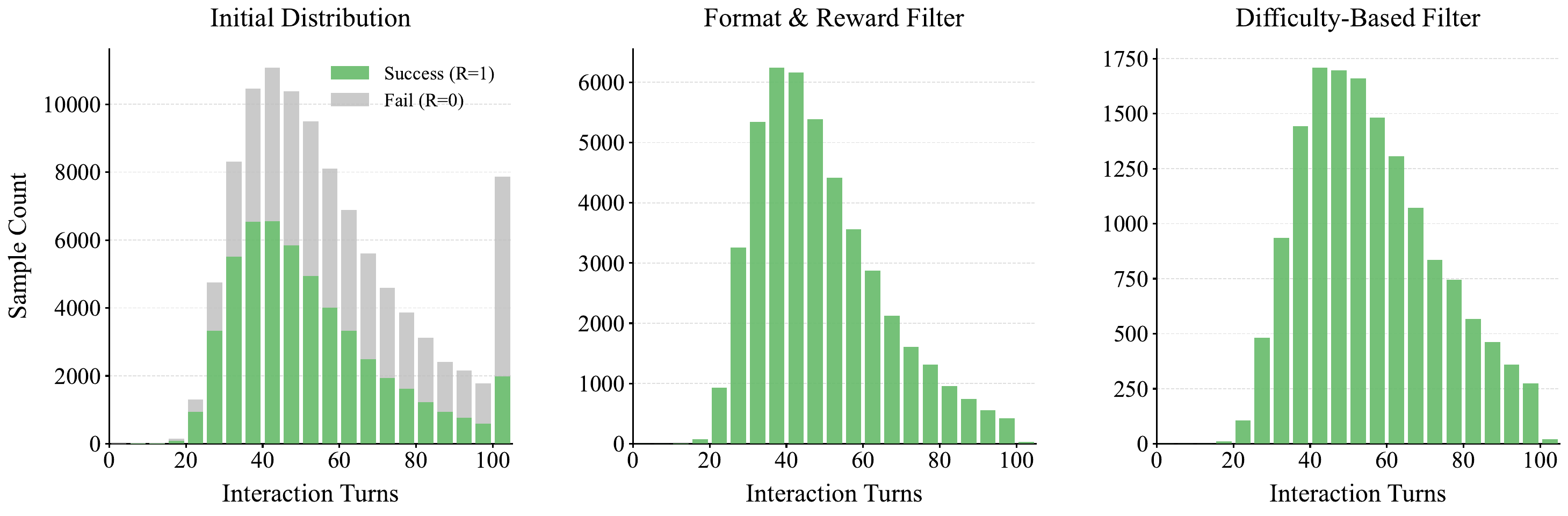}
\caption{Evolution of interaction turn distributions through sequential filtering stages. From left to right: (1) the initial distribution of successful and failed trajectories; (2) the distribution after applying format and reward constraints; and (3) the final refined distribution following difficulty-based filtering.}
    \label{fig:data_filter_dist}
\end{figure}

Figure~\ref{fig:data_filter_dist} shows the evolution of the trajectory length distribution across the format-based and difficulty-based filtering stages. The results show that the filtering pipeline effectively removes outliers, particularly long-tail failure cases present in the initial distribution, leading to a substantially smoother and more stable distribution of trajectory lengths in the final SFT dataset.

\subsection{Long-Horizon Supervised Fine-Tuning}
Based on the filtered dataset and the corresponding trajectories, we perform multi-turn SFT on the Qwen2.5-Coder-32B-Instruct~\citep{hui2024qwen2} and Qwen3-4B-Instruct-2507 models~\citep{yang2025qwen3technicalreport}. We apply YaRN~\citep{peng2023yarnefficientcontextwindow} to extend the maximum context length from 32K to 80K tokens, enabling effective modeling of long multi-turn trajectories. During training, we adopt a multi-turn masking strategy that excludes environment feedback obtained from Docker-based execution from the loss computation, ensuring that the model focuses on learning reasoning and action generation rather than fitting execution outputs. 
After training on approximately 60K instances, we obtain \textit{SWE-Master-SFT}, which serves as the initialization for the subsequent RL training. A detailed analysis of data scaling is provided in Section~\ref{sec:ana:sft_data_scaling}, while the impact of data filtering is discussed in Section~\ref{sec:ana:data_filter}.

\subsection{Reinforcement Learning with Real Environments}

This section provides an overview of the RL stage built upon SWE-Master. The data distribution used for RL is aligned with that of the SFT dataset, ensuring consistency between the two training phases. All interactions during training are conducted within real Docker-based execution environments, which guarantees the reliability and fidelity of environmental feedback.

\subsubsection{Policy Optimization Algorithm}

We adopt Reinforcement Learning with Verifiable Reward (RLVR) as our foundational paradigm, employing the Group Relative Policy Optimization (GRPO) algorithm~\citep{shao2024deepseekmath} to optimize \textit{SWE-Master-SFT} directly. Building upon the established efficacy of prior reinforcement learning methodologies~\citep{wang2025reinforcement, liu2025understanding, deepswe2025, yu2025dapo}, we incorporate several empirical optimizations to enhance training stability and performance:

\paratitle{Leave-One-Out Advantage Estimation.} To reduce the variance of the policy gradient estimate without introducing bias, we compute the advantage for each sample by normalizing its reward against the average reward of the other samples in the group, excluding the sample itself.

\paratitle{Mitigation of Inherent Bias.} 
To prevent biased optimization, we modify two standard normalization terms. 
We replace the dynamic division by the trajectory length $1/|o_i|$ (where $o_i$ represents the $i$-th generated trajectory for a prompt) with a fixed constant scaling, as formulated in Eq.~\ref{eq:grpo}.
This modification prevents the policy from favoring brevity in correct answers or verbosity in incorrect ones. 
Additionally, we omit the standard deviation normalization from the advantage calculation to avoid biasing updates based on task difficulty variance.

\paratitle{Clip-Higher.} To counter the phenomenon of entropy collapse and sustain exploration, we adopt the clip-higher strategy. This modification relaxes the upper clipping bound, addressing the limitation in standard GRPO where the probability growth of low-likelihood ``exploration'' tokens is disproportionately constrained compared to high-likelihood ``exploitation'' tokens.

\paratitle{Removal of KL Divergence.} We eliminate the KL divergence penalty from the objective function. This unbinds the policy from the trust region of the initial SFT reference model, granting the model the flexibility to optimize more aggressively towards the verifiable reward signal.

Incorporating these modifications, the final objective function is defined as follows:

\begingroup
\small
\begin{equation}
\label{eq:grpo}
\mathcal{J}(\theta) = \mathbb{E}_{q \sim P(Q), \{o_i\}_{i=1}^G \sim \pi_{\text{old}}} \left[ \frac{1}{G} \sum_{i=1}^G \frac{1}{L_{\text{max}}} \sum_{t=1}^{|o_i|} \min \left( \rho_{i,t}(\theta) \hat{A}_i, \text{clip}\left(\rho_{i,t}(\theta), \beta_{\text{low}}, \beta_{\text{high}}\right) \hat{A}_i \right) \right]
\end{equation}
\endgroup
where $P(Q)$ denotes the distribution of input prompts, and the clipping bounds are defined as $\beta_{\text{low}} = 1-\varepsilon_{\text{low}}$ and $\beta_{\text{high}} = 1+\varepsilon_{\text{high}}$. The term $L_{\text{max}}$ represents a fixed constant for length normalization. The probability ratio $\rho_{i,t}(\theta)$ and the group-relative advantage $\hat{A}_i$ are given by:
\begingroup \small
\begin{equation*}
\rho_{i,t}(\theta) = \frac{\pi_\theta(o_{i,t}|q, o_{i,<t})}{\pi_{\text{old}}(o_{i,t}|q, o_{i,<t})}, \quad \hat{A}_i = R_i - \frac{1}{G-1}\sum_{j=1, j\neq i}^G R_j.
\end{equation*}
\endgroup

\subsubsection{Reward Design}

We employ a standard binary outcome-based reward function aligned with the evaluation protocols of SWE-bench. For a given issue, if the patch submitted by the policy model successfully passes all F2P and P2P unit tests, the reward is $r=1$; otherwise, the reward is $r=0$.

Although SWE-Master is fine-tuned on long-context trajectories (up to 80K tokens) during the SFT phase, extending this capability to RL presents significant challenges. We observe that during the initial stages of RL, the policy model sometimes exhibits uncertainty, engaging in repetitive cycles of unit test generation and modification without issuing a final submission. This behavior leads to a high rate of trajectory truncation (approximately 20\%) due to exhaustion of token or turn budgets. Consequently, the model receives zero rewards despite making partial progress, which fails to reinforce effective reasoning behaviors and causes the average training reward to degrade (see Figure~\ref{fig:rl_dynamics}), eventually leading to training collapse. However, offline evaluation of the patches generated during the RL process reveals that approximately 24.3\% of the truncated trajectories—those without a final submission—successfully resolve the current issue. This failure to submit primarily stems from model underconfidence, which drives the agent into redundant verification loops, or from the generation of excessively rigorous unit tests that create self-imposed barriers to submission.

To address this pathology and stabilize training, we implement a \textit{forced submission mechanism}, coupled with a reward shaping strategy. 
Specifically, if a trajectory terminates due to budget exhaustion (\eg timeout or maximum turns) rather than a voluntary submission, we enforce a patch submission based on the current state of the repository to evaluate its correctness.  This mechanism ensures that each trajectory yields a concrete submission.
We then apply a stop-reason-dependent modulation to the final reward.  

Furthermore, the training process relies on launching and executing containerized
environments. Due to hardware or storage-related instability on machines hosting Docker images,
container startup failures may occasionally occur. Such failures are unrelated to the policy model’s
behavior and can introduce spurious noise into training if not properly handled. To preserve training
stability and integrity, we apply a \textit{container error masking} strategy: all tasks affected by container startup
failures are masked and excluded from reward computation and policy updates. This design prevents
infrastructure-level errors from negatively influencing the learning process. Based on the design principles discussed above, our reward design is formulated as follows:

\begin{align}
    R &= 
    \begin{cases} 
        r_{\text{outcome}} & \text{if } s = \texttt{DONE} \\
        \alpha \cdot r_{\text{outcome}} & \text{if } s \in \{\texttt{TIMEOUT}, \texttt{MAX\_STEPS}, \texttt{MAX\_TOKENS}\} \\
        0 & \text{if } s \in \{\texttt{CONTAINER\_FAILED}\}
    \end{cases} \\[1ex] 
    M &= 
    \begin{cases} 
        0 & \text{if } s \in \{\texttt{CONTAINER\_FAILED}, \texttt{SERVER\_ERROR}\} \\
        1 & \text{otherwise}
    \end{cases}
\end{align}

where $R$ and $M$ denote the final shaped reward and the loss mask, respectively. Here, $r_{\text{outcome}} \in \{0, 1\}$ represents the binary evaluation result of the patch, and $s$ signifies the trajectory termination reason. The parameter $\alpha \in (0, 1)$ is a penalty coefficient for forced submissions, set to $0.5$ in our experiments. The specific categories of termination conditions $s$ are defined as follows:

\begin{itemize}[leftmargin=2.0em]
\item \texttt{DONE} represents the agent voluntarily  submit a patch  before reaching resource limits.
\item \texttt{TIMEOUT}, \texttt{MAX\_STEPS}, and \texttt{MAX\_TOKENS} 
denote forced terminations due to budget exhaustion (\eg reaching temporal, token, or interaction step limits).
\item \texttt{CONTAINER\_FAILED} refers to unforeseen infrastructure-level failures (\eg hardware issues or environment crashes).
\end{itemize}

This design serves multiple strategic objectives.  Primarily, it ensures training integrity by masking loss of failed trajectories stemming from container runtime errors, guaranteeing that policy updates are driven exclusively by valid agent-environment interactions. Meanwhile, the modulated reward coefficient $\alpha$ calibrates the trade-off between exploration and efficiency; it encourages the model to undertake the extended reasoning necessary for complex tasks while implicitly penalizing redundant behaviors that lead to timeout. Finally, the forced submission mechanism mitigates reward sparsity by validating potential solutions even upon budget exhaustion, thereby preventing the training collapse associated with vanishing signals—a benefit further substantiated in Section~\ref{sec:ana:reward_design}.

\subsubsection{Training Tricks}

\paragraph{Budget Awareness.}
Prior studies~\citep{liu2025budget} demonstrate that explicitly providing agents with budget-related signals during long-horizon interactions enables more rational behavior planning and more effective allocation of limited interaction steps. Motivated by these findings, we incorporate \emph{budget awareness} into the agent--environment interaction process. Concretely, at the end of each interaction turn, the environment returns not only the standard feedback but also the remaining budget, defined as the number of turns left before termination. This design allows the policy model to explicitly condition its decisions on the remaining interaction budget, thereby encouraging more deliberate planning and earlier convergence toward a viable solution. The corresponding environment additional repsonse is as follows:


\begin{PromptBoxNew}[p:budget_awareness]{Environment Response for Budget Awareness}
\small
This is step STEP\_INDEX of a maximum of MAX\_STEPS.  Steps Remaining: STEP\_REMAINING.
\end{PromptBoxNew}

\paragraph{Restriction of Git Commands.}
Consistent with the trajectory distillation stage in SFT, we strictly restrict the use of \texttt{git}-related commands during reinforcement learning. In particular, when the model attempts to invoke commands such as \texttt{git log} or \texttt{git show}, which may reveal solution-relevant information without genuine reasoning, the environment immediately returns a warning message. The model is explicitly informed that such behavior is prohibited and is encouraged to rely on its own analysis and code modifications instead. This constraint is designed to prevent shortcut exploitation and to ensure that the learned policy reflects authentic problem-solving capabilities. The corresponding environment repsonse is as follows:
\begin{PromptBoxNew}[p:git_ban_response]{Environment Response for Git Command}
\small
Bash command 'git show' and 'git log' is not allowed. Please use a different command or tool.
\end{PromptBoxNew}

\paragraph{Environmental Response Masking.} 
During training, we implement environment response masking to exclude environment feedback from both loss and advantage calculations. This strategy ensures that only the tokens generated by the model itself contribute to the optimization process, thereby enhancing training stability and preserving the structural integrity of the agent’s reasoning sequences.

\subsubsection{Dynamics of Policy Learning in RL}

\begin{figure}[ht]
\centering
    \centering
    \includegraphics[width=1.0\linewidth]{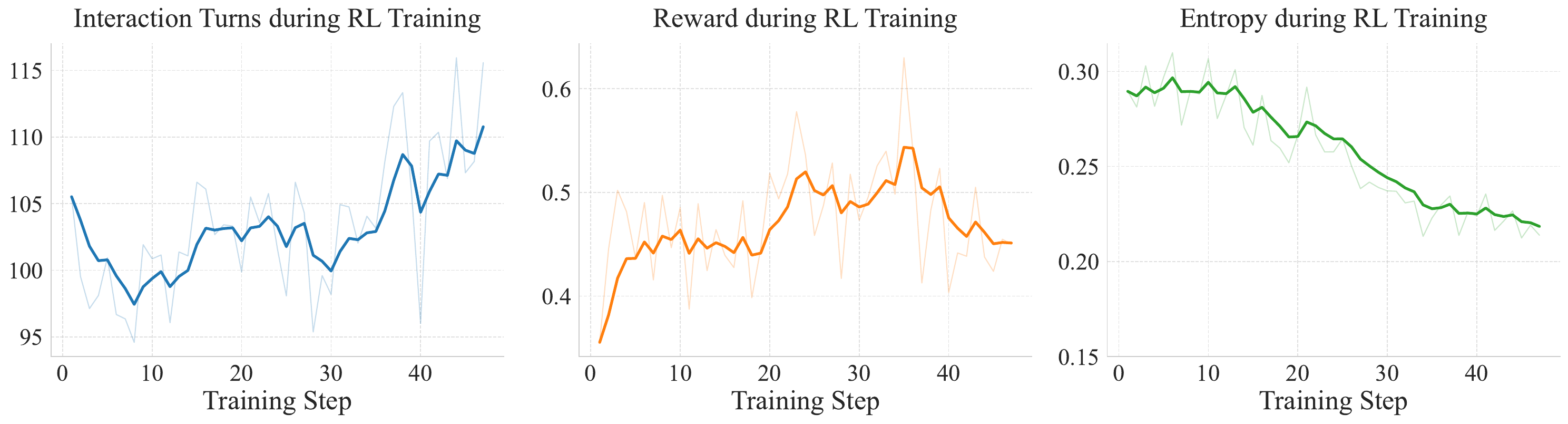}
\caption{ The training dynamics of interaction turns, reward and entropy for SWE-Master RL training.}
    \label{fig:rl_dynamics}
\end{figure}

Figure~\ref{fig:rl_dynamics} illustrates the RL training dynamics of the policy model, where a clear synergy between performance gains and behavioral adaptation emerges. 
The \textbf{reward} exhibits a consistent upward trajectory from an initial value of 0.35 toward a stabilized peak, demonstrating that the model effectively learns to navigate repository environments to secure verifiable rewards. 
This performance improvement is accompanied by a gradual increase in \textbf{interaction turns}, suggesting that as the agent matures, it learns to utilize a larger interaction budget—likely for more exhaustive exploration and rigorous self-verification—to tackle complex software issues. 
Simultaneously, the \textbf{entropy} follows a steady decline without exhibiting the phenomenon of entropy collapse.
Together, these metrics signify a stable and effective reinforcement learning process.

\subsection{Test-Time Scaling}
Test-Time scaling (TTS) significantly enhances the performance of SWE agents  by leveraging increased inference-time compute to navigate complex task spaces~\citep{chen2024expanding, snell2024scaling}.
Typically, TTS is implemented via two primary paradigms~\citep{tao2026swe}: sequential scaling, which involves increasing the allowance of interaction turns; and parallel scaling, which generates multiple trajectories and corresponding patches, followed by employing a specific verifier to select the optimal candidate for submission. In this section, we investigate the performance of SWE-Master under both scaling strategies.

\subsubsection{Sequential Scaling}
\begin{figure}[ht]
\centering
    \centering
    \includegraphics[width=0.9\linewidth]{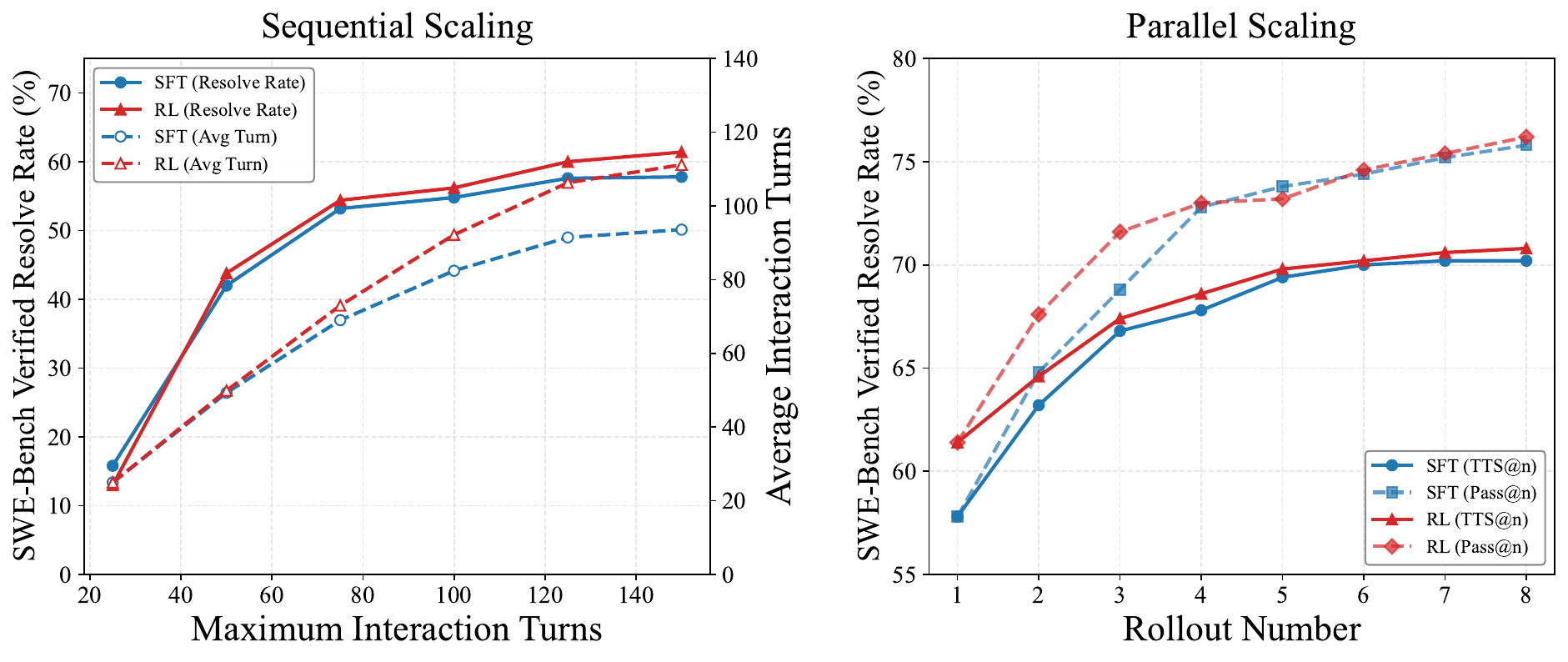}
\caption{Performance scaling of SWE-Master via test-time scaling. \textbf{Left:} Sequential scaling illustrates the impact of increasing interaction turn limits. \textbf{Right:} Parallel scaling demonstrates the performance gains achieved through increased rollout breadth and candidate verification.}
    \label{fig:tts_scaling}
\end{figure}

Sequential scaling is implemented by extending the maximum limits on generated tokens and interaction turns. With this expanded computational budget, the model is empowered to explore the repository structure more comprehensively and generate more unit tests to validate its proposed modifications, thereby enhancing the accuracy of the final submitted patch.
Given the significant heterogeneity in file counts and repository content across different problem instances, which leads to substantial variance in the token consumption for environment feedback and code modifications, we adopt the number of interaction turns as the primary constraint metric rather than token consumption. We fix the maximum context length at 128K to isolate the impact of interaction depth, and systematically scale the maximum interaction limit from 25 to 150.

As illustrated in Figure~\ref{fig:tts_scaling}, increasing the interaction budget from 25 to 150 turns yields significant performance gains for both models. The SWE-Master-RL model demonstrates superior scalability, effectively leveraging the extended turn limit to a 61.4\% resolve rate, consistently outperforming the SWE-Master-SFT model. While average turn usage (dashed lines) indicates active utilization of the available budget, the resolve rates for both models begin to plateau beyond 125 turns, suggesting diminishing marginal returns at higher limits.
\subsubsection{Parallel Scaling}

Parallel scaling enhances problem-solving performance by generating multiple candidate trajectories for a single issue and employing a selection mechanism to identify the most viable patch. The efficacy of this approach hinges critically on the ability to distinguish correct solutions from incorrect ones within a pool of candidates.
Existing approaches~\citep{shum2025swe, swe-gym} typically rely on training a verifier to predict a scalar probability (\ie a binary Yes/No classification) based on the generation trajectory or the final patch. However, these black-box verifiers often lack interpretability and struggle to accurately assess correctness when verifying long, noisy context windows characteristic of repository-level software engineering. To address these limitations, we use the \textbf{SWE-World} model~\citep{swe-world}, a specialized model designed to perform simulated evaluation rather than simple probability estimation.

Unlike traditional verifiers, SWE-World aligns the selection process with the formal execution environment found in SWE-bench tasks. Upon the completion of a candidate trajectory, SWE-World extracts the context relevant to the evaluation, including the modified files, test cases, and the patch itself. It then processes this input to generate a simulated test report and a predicted reward. This approach provides a granular, interpretable rationale for selection, mimicking the feedback of a real compiler and test runner.

We construct the training data for SWE-World through an offline rollout process. For every patch applied during the rollout, we capture the relevant execution context alongside the ground-truth test report and the actual reward obtained from the environment. To ensure high data quality, we employ Qwen3-235B-A22B-Thinking-2507 to perform reverse reasoning~\citep{lin2025scaling}. This teacher model analyzes the causal relationships between the input context and the structured evaluation feedback to reconstruct reasoning chains, thereby creating a  logically coherent dataset for SFT.

To implement parallel scaling, we generate $N$ independent rollout trajectories for a given issue using the policy model, yielding a candidate trajectory set $\mathcal{T} = \{ \tau_1, \tau_2, \dots, \tau_n\}$ and a corresponding patch set $\mathcal{P} = \{p_1, p_2, \dots, p_N\}$. Subsequently, to robustly estimate the correctness of each solution, we perform $K$ stochastic reward simulations for each trajectory $\tau_i$ using the reward model (setting $K=3$ in our experiments). Formally, the optimal trajectory $\tau^*$ is identified by maximizing the expected reward, which is approximated by the arithmetic mean of the sampled rewards:


\begin{equation}
    \tau^* = \operatorname*{argmax}_{\tau_i \in \mathcal{T}} \left( \frac{1}{K} \sum_{j=1}^K \hat{r}_{i,j} \right),
\end{equation}

where $\hat{r}_{i,j}$ denotes the reward obtained from the $j$-th simulation iteration for candidate $\tau_i$. In cases where multiple candidates achieve the maximal score, ties are broken via random selection. Finally, the patch $P^*$ associated with the selected trajectory $\tau^*$ is submitted to the environment as the definitive solution for the current issue.

As illustrated in Figure~\ref{fig:tts_scaling}, increasing the number of rollouts yields a consistent improvement in the SWE-bench Verified resolve rate for both the SFT and RL models. {Pass@$K$} (dashed lines) represents the theoretical upper bound where an oracle selects the correct patch, while {TTS@$K$} (solid lines) reflects the actual performance using the SWE-World. We observe that the RL model consistently outperforms the SFT baseline, starting at 61.4\% and surpassing 70.8\% at TTS@8. Crucially,TTS@$K$ performance closely tracks the theoretical optimal curves (Pass@$K$), indicating that our selection mechanism is highly effective at identifying the correct solution within the generated candidate pool. This strong alignment validates that the simulated evaluation approach successfully converts the increased computational budget into tangible performance gains, minimizing the gap between potential and realized accuracy.

To evaluate the fidelity of the reward predictions, we assess the alignment between the simulated rewards and the ground-truth environmental feedback across the evaluated trajectories. As presented in Table~\ref{tab:dockerllm_performance}, our trained SWE-World achieves an accuracy of 77.59\%, with a recall of 71.40\% and a precision of 71.64\%. This performance confirms that the SWE-World effectively functions as a high-fidelity simulated execution environment, enabling it to serve as a dependable surrogate for the actual sandbox during candidate patch selection.

\begin{table}[ht]
    \centering
    \small
    \caption{Comparison of Precision, Recall, and Accuracy for trajectory reward prediction. Real-Docker serves as the ground truth baseline for evaluating SWE-World.}
    \label{tab:dockerllm_performance}
    \setlength{\tabcolsep}{12pt} 
    \begin{tabular}{cccc}
        \toprule
        \textbf{Method} & \textbf{Precision} & \textbf{Recall} & \textbf{Accuracy} \\
        \midrule
        Real-Docker & 100.00 & 100.00 & 100.00 \\
        SWE-World & 77.59 & 71.40 & 71.64 \\
        \bottomrule
    \end{tabular}
\end{table}

\section{Experiment}
\label{exp}

\subsection{Experimental Settings}
\paratitle{Evaluation Datasets.}
We primarily evaluate our method on the SWE-bench Verified~\citep{sweb-verified} dataset, which consists of 500 solvable instances curated from the original SWE-bench benchmark~\citep{jimenez2024swebench}. The dataset is constructed from real-world GitHub issues spanning 12 open-source python repositories and is designed to assess the end-to-end issue resolution capabilities of LLMs. The verified split provides a more reliable evaluation setting by restricting instances to those with confirmed valid solutions.

\paratitle{Evaluation Metrics.}
The model's performance is quantified by the resolve rate, representing the percentage of tasks successfully addressed, following the protocol detailed in Section~\ref{sec:preliminaries_verify}.

\paratitle{Baselines.}
We employ Qwen2.5-32B-Coder-Instruct~\citep{hui2024qwen2} and Qwen3-4B-Instruct-2507~\citep{yang2025qwen3technicalreport} as backbone models. To evaluate SWE-Master, we compare its performance against leading open-source code agents~\citep{hui2024qwen2, yang2025qwen3technicalreport, swe-gym, jain2025r2e, zeng2025skywork, yang2025swesmith, cao2025skyrl, SWESwiss2025, deepswe2025, wang2025swe, yang2025kimi, sonwane2025bugpilot, liu2025context, tao2026swe, zeng2026davinci, seed2025seed-oss, copet2025cwm, xie2025swe}, and frontier open-source foundations~\citep{minimax_m21, glm_47, kimiteam2025kimik2openagentic, liu2025deepseek, xiao2026mimo, agarwal2025gpt}. Given that the design of an agent's scaffold significantly influences final resolve rates, we directly cite the metrics reported in the original publications to ensure a fair comparison under each model's intended optimal configuration.

\paratitle{Implementation Details.}\label{sec:impele_details}
We utilize R2E-Gym framework, a streamlined adaptation of the OpenHands, for trajectory generation, while leveraging OpenRLHF and RLLM for SFT and RL, respectively. During the SFT phase, models are trained over 5 epochs using a maximum context length of 80K tokens and a global batch size of 256. We employ a cosine learning rate scheduler, decaying from a peak of $5\times10^{-5}$ to $5\times10^{-6}$, with a warmup ratio of 0.1. Subsequently, in the RL stage, we utilize a constant learning rate of $1\times10^{-6}$ and a batch size of 32 problems, with each problem generating 4 parallel rollouts at a sampling temperature of 1.0. The exploration is constrained by a per-trajectory timeout of 5,400 seconds, a maximum interaction turn of 150 turns, and a context window of 108K tokens. For final evaluation, inference is performed with a temperature of 0.7, a maximum context capacity of 128K tokens, and the maximum interaction turn is 150.

\subsection{Main Results}

\begin{table}[htbp]
    \small
    \centering
    \setlength{\tabcolsep}{6pt} 
    \begin{tabular}{lcccc}
        \toprule
        \textbf{Model/Method} &\textbf{BackBone}& \textbf{Scaffold} & \textbf{Training} & \textbf{Resolve Rate (\%)} \\
        \midrule
        \multicolumn{5}{c}{\textbf{Open-Source Foundation Models }} \\
        \midrule
        Minimax-M2.1 &-& Internal & - & 74.0 \\
        GLM-4.7 &-&Internal  & - & 73.8 \\
        DeepSeek-V3.2 &- & Internal &-  & 73.1\\
        GPT-OSS-20B&-& Internal &- & 60.7\\
        GPT-OSS-120B&-& Internal &- & 62.4\\
        \midrule
        \multicolumn{5}{c}{\textbf{Open-Source Code Agents}} \\
        \midrule
        Qwen2.5-Coder-32B  &- & OpenHands & - & 6.2 \\
        Qwen3-Coder-30B-A3B& - & OpenHands & - & 51.6 \\
        SWE-Gym-32B& Qwen2.5-Coder-32B-Inst & OpenHands & SFT & 20.6 \\
        R2E-Gym-32B & Qwen2.5-Coder-32B-Inst  & R2E-Gym & SFT & 34.4 \\
        \quad + TTS@16 &  Qwen2.5-Coder-32B-Inst  & R2E-Gym & SFT & 49.4 \\
        Skywork-SWE-32B & Qwen2.5-Coder-32B-Inst & OpenHands & SFT & 38.0 \\
        \quad + TTS@8 & Qwen2.5-Coder-32B-Inst & OpenHands & SFT & 47.0 \\
        SWE-Fixer-72B & Qwen2.5-72B-Base & Agentless & SFT &32.8 \\
        SWE-agent-LM-32B& Qwen2.5-Coder-32B-Inst & SWE-agent & SFT & 40.2 \\
        SA-SWE-32B  & Qwen3-32B & OpenHands & RL & 39.4 \\
        SWE-Swiss-32B & Qwen2.5-32B-Inst & Agentless & SFT+RL &58.0 \\
        DeepSWE-32B-Preview & Qwen3-32B & OpenHands & RL & 42.2 \\
        \quad + TTS@16 & Qwen3-32B & OpenHands & RL & 59.0 \\
        Seed-OSS-36B & - & OpenHands & -  & 56.0 \\
        SWE-Mirror-LM-32B & Qwen2.5-32B-Inst & MOpenHands & SFT & 52.2 \\
        Kimi-Dev-72B& Qwen2.5-72B-Base  & SWE-Agent  &SFT+RL & 48.6 \\
        \quad + TTS@40 & Qwen2.5-72B-Base & Agentless & SFT+RL & 60.4 \\
        FrogBoss-32B& Qwen3-32B  & SWE-Agent  & SFT+RL & 54.6 \\
        SWE-Compressor& Qwen2.5-Coder-32B-Inst &OpenHands & SFT & 57.6\\
        SWE-Lego-Qwen3-32B& Qwen3-32B & OpenHands & SFT & 52.6 \\
        \quad + TTS@16& Qwen3-32B  & OpenHands & SFT & 58.8 \\

        daVinci-Dev-32B& Qwen2.5-32B-Base  & SWE-Agent  & MT+SFT & 56.1 \\
        daVinci-Dev-72B& Qwen2.5-72B-Base  & SWE-Agent &MT+SFT & \underline{58.5} \\
        
        \midrule
        \textbf{SWE-Master-4B-SFT}&Qwen3-4B-Inst-2507 & R2E-Gym & SFT & 27.6 \\
        \textbf{SWE-Master-4B-RL}&Qwen3-4B-Inst-2507 & R2E-Gym & SFT+RL & 33.4 \\
        \textbf{SWE-Master-32B-SFT}& Qwen2.5-Coder-32B-Inst & R2E-Gym & SFT & 57.8  \\
        \quad + TTS@8 &Qwen2.5-Coder-32B& R2E-Gym & SFT+RL & 70.2 \\
        \textbf{SWE-Master-32B-RL} & Qwen2.5-Coder-32B-Inst & R2E-Gym & SFT+RL & \textbf{61.4} \\
        \quad + TTS@8&Qwen2.5-Coder-32B & R2E-Gym & SFT+RL & 70.8 \\
        \bottomrule
    \end{tabular}
\caption{Performance comparisons between SWE-Master and the baselines on SWE-bench Verified. The baseline results are referenced from their respective papers. The best and second best {Pass@1} results are highlighted in  \textbf{bold} and \underline{underlined}, respectively.}

\label{tab:main_results} 
\end{table}

Table\ref{tab:main_results} shows the results of SWE-Master and the baselines on SWE-bench verified. We can obtain the following observations: 

\noindent $\bullet$ \emph{Achieving Frontier Performance among Open-Source Code Agents.}
SWE-Master demonstrates superior performance compared to existing open-source code agents. Specifically, our SWE-Master-32B-RL model achieves a resolve rate of 61.4\% at Pass@1, significantly outperforming strong baselines such as daVinci-Dev (58.5\%) and SWE-SWE-Compressor (57.6\%). This indicates that our framework effectively unleashes the potential of the 32B parameter models, achieving high efficacy without relying on excessive model scaling.

\noindent $\bullet$ \emph{Demonstrating the Efficacy of Reinforcement Learning across Scales.} 
The transition from SFT to RL consistently yields notable performance gains across different model scales, validating the robustness of our RL framework. For the smaller Qwen3-4B-2507 backbone, RL training boosts the resolve rate from 27.6\% to 33.4\% (an absolute improvement of 5.8\%). Similarly, on the larger Qwen2.5-Coder-32B backbone, the performance increases from 57.8\% to 61.4\%. These results confirm that our RL strategy effectively guides models to explore and internalize complex verification strategies beyond simple behavior cloning.

\noindent $\bullet$ \emph{Unlocking Peak Performance via Test-Time Scaling.} 
Incorporating Test-Time Scaling via the simulated verification and ranking mechanism using SWE-World, yields substantial improvements in resolve rates. With a moderate compute budget of 8 rollouts (TTS@8), SWE-Master-32B-SFT improves by 12.4\% (from 57.8\% to 70.2\%), and SWE-Master-32B-RL improves by 9.4\% (from 61.4\% to 70.8\%). Notably, our TTS@8 approach is significantly more efficient and effective than competitors utilizing larger budgets, such as DeepSWE-32B + TTS@16 (59.0\%) and R2E-Gym-32B + TTS@16 (49.4\%). These results demonstrate the effectiveness of combining a strong policy with a robust selection mechanism.

\subsection{Observations on Model Behavior during Evaluation}

\begin{figure}[ht]
\centering
    \centering
    \includegraphics[width=0.8\linewidth]{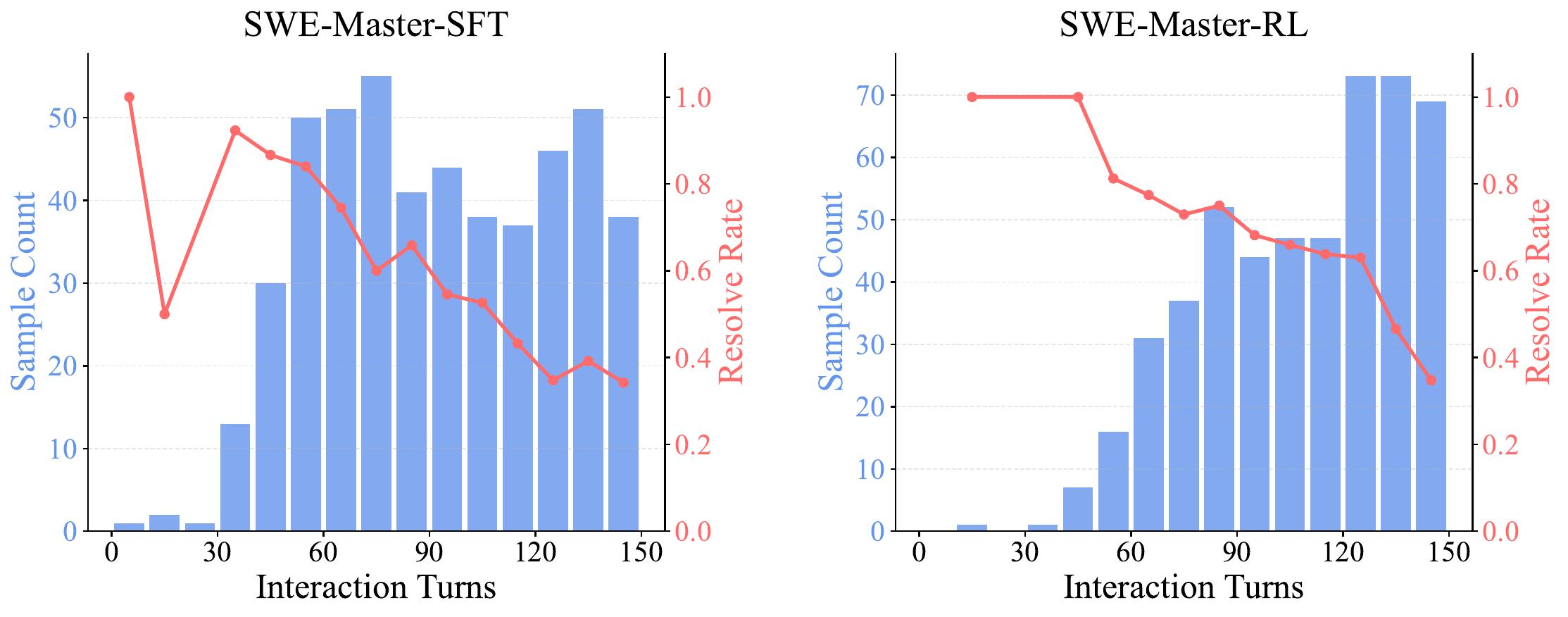}
\caption{Resolve rates and sample distributions of SWE-Master with respect to the number of interaction turns, based on the corresponding evaluation results.}
    \label{fig:turn_passrate_distribution}
\end{figure}

\begin{figure}[ht]
\centering
    \centering
    \includegraphics[width=0.8\linewidth]{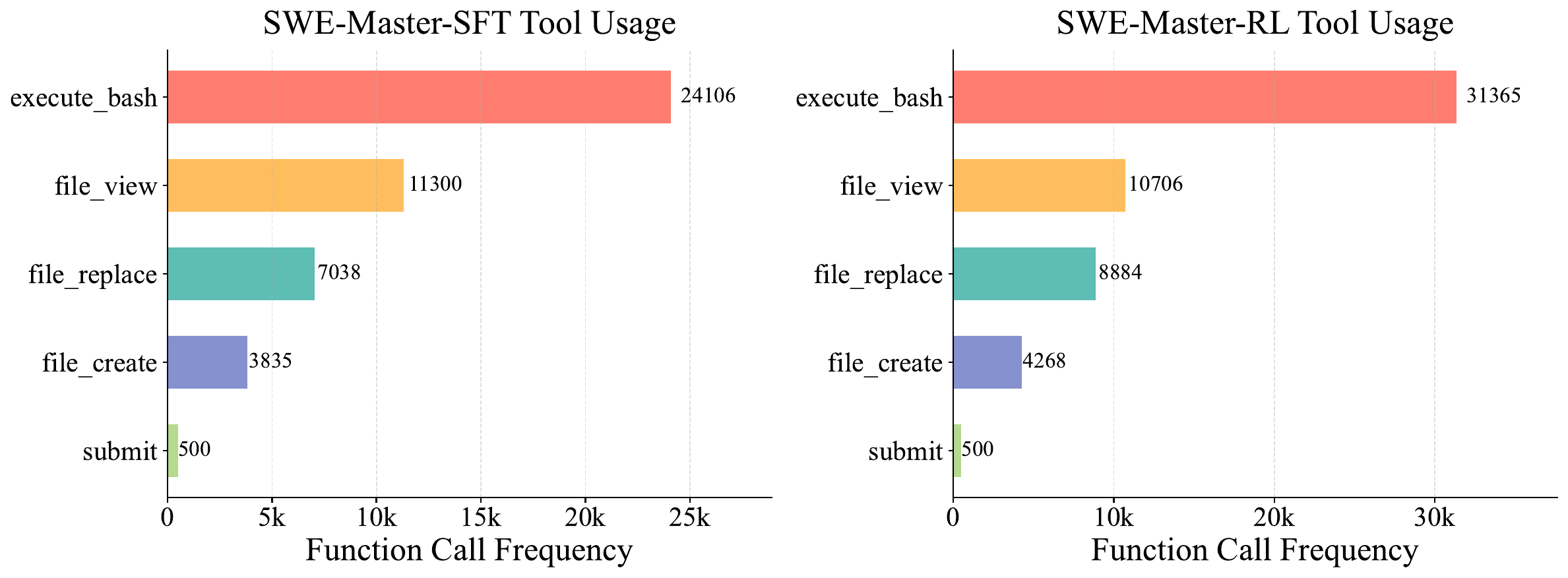}
\caption{Distribution of tool usage in the evaluation results of SWE-Master.}
    \label{fig:tool_usage_distribution}
\end{figure}

In this section, we analyze the behavior exhibited by SWE-Master during the evaluation. As illustrated in Figure~\ref{fig:turn_passrate_distribution}, we observe a general inverse correlation between trajectory length and resolve rate for both models, confirming that instances requiring extended reasoning chains are inherently more prone to failure due to the higher intrinsic complexity of these tasks. Despite this difficulty, the RL model exhibits a pronounced distributional shift toward the upper limits of the turn budget (peaking at 120--140 turns), contrasting with the SFT model's tendency to submit earlier. This varying interaction turn is intrinsically linked to the tool usage patterns shown in Figure~\ref{fig:tool_usage_distribution}, where the frequency of \texttt{execute\_bash} and \texttt{file\_editor\_replace} operations increases significantly in the RL model. This suggests that the reinforcement learning process encourages the agent to engage in more active iterative debugging—leveraging the expanded budget to repeatedly execute tests and modify code.

\section{IDE-Level Code Capacity}
\label{sec:ide_injection}


This section introduces the integration of \texttt{lsp\_tool}, which empowers the code agent with a holistic understanding of the codebase and facilitates precise navigation of complex repository structures. We systematically evaluate its impact across training and inference phases, showcasing how this design bridges the gap between basic bash-level execution behavior and senior-level capability.

\subsection{From Grep to LSP-Based Code Navigation}

Current frontier software agent frameworks, such as OpenHands~\cite{wang2025openhands} and SWE-agent~\cite{yang2024swe} encounter significant bottlenecks when addressing non-crashing defects where the context is ambiguous. Predominantly, these systems rely on Linux CLI-based lexical stream retrieval tools (\eg \texttt{grep}, \texttt{find}) for code context localization. While effective for simple pattern matching, these tools lack semantic understanding. When identifying bugs involving heavily overloaded function names or multi-level function calls across files and repositories, such methods~\cite{wang2025openhands, yang2024swe, jain2025r2e} are not only inefficient and unreliable but often retrieves irrelevant contexts that obscure true semantic correlations and call relationships, severely hindering LLMs from achieving robust defect repair. While some studies have recognized these limitations and incorporated Abstract Syntax Tree (AST) for localization~\cite{zhang2025one,dong2025infcode,jiang2025cosil}, they lack standardized interfaces, cross-language scalability, and rigorous verification on benchmarks like SWE-bench Verified.

To bridge this semantic gap, we propose a novel approach inspired by modern integrated development environments (IDEs). We introduce the first unified, language-agnostic code navigation tool for agents based on the \textbf{Language Server Protocol (LSP)}. In contrast to previous ad-hoc approaches, our \texttt{lsp\_tool} implementation provides a standardized interface that enables advanced semantic features—including precise \textit{go to definition} and \textit{find references}—consistent with modern IDEs. We integrated this tool into the framework described in Section~\ref{sec:framework_and_env} and initially validated its effectiveness using powerful open-source foundation models, specifically MiniMax-M2.1 and GLM-4.7, confirming that \texttt{lsp\_tool} significantly enhances repository-level understanding. Building on this validation, we employ \textit{distillation} to endow our SWE-Master model with this advanced IDE-level code navigation capability, enabling it to master complex repository exploration with the efficiency required for real-world deployment.

\subsection{Overview of LSP Tools}

In this section, we introduce the foundational technology behind our approach, detail the implementation of the \texttt{lsp\_tool} specifically designed for code agents, and illustrate how this tool is integrated into the agentic workflow.

\subsubsection{The Language Server Protocol}

\begin{figure}[htbp] 
    \centering
    \begin{subfigure}[b]{0.29\textwidth} 
        \centering
        \includegraphics[width=\linewidth]{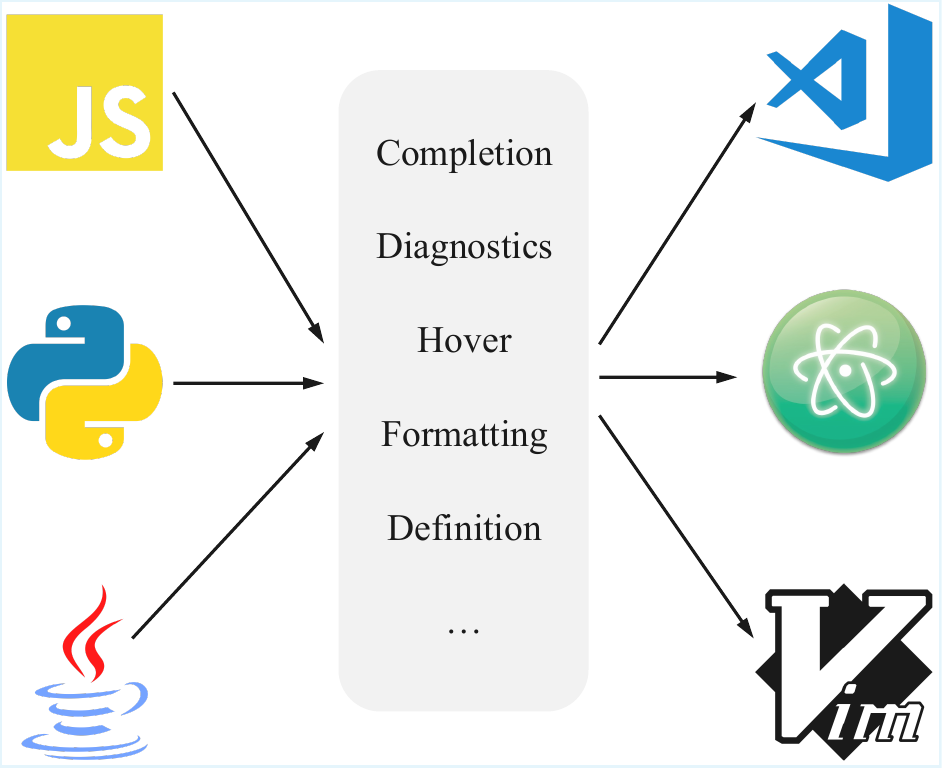} 
        \caption{Language Server Protocol}
        \label{fig:lsp-language-editors-main}
    \end{subfigure}
    \hfill 
    \begin{subfigure}[b]{0.68\textwidth}
        \centering
        \includegraphics[width=\linewidth]{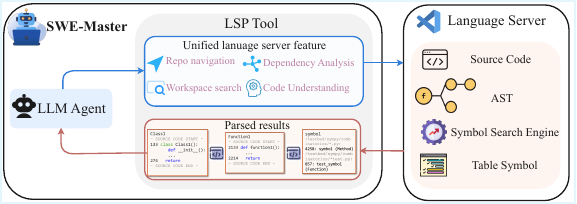}
        \caption{Illustration of LLM navigating through a code repository with LSP tools.}
        \label{fig:lsp-pipline}
    \end{subfigure}
    
    \caption{Language Server Protocol from supporting integrated development environments to supporting code agents navigating through a code repository. }
    \label{fig:lsp_main_figure}
\end{figure}

The Language Server Protocol~\cite{lsp}, originally introduced by Microsoft, is a pivotal standard designed to unify the interaction between development tools (clients) and language-specific intelligence providers (servers). It defines a standardized communication protocol based on JSON-RPC that decouples the IDE from the underlying language compilers or analysis engines. This decoupling allows text editors and IDEs to interact with different programming languages through a uniform interface, eliminating the need for $M$-to-$N$ point-to-point integrations, as depicted in Figure~\ref{fig:lsp-language-editors-main}. Appendix~\ref{details of lsp Protocol} shows more details about this.

Taking the popular IDE—Visual Studio Code (VSCode) as a prime example, LSP provides a unified interface that facilitates seamless communication between the editor and language servers. This integration empowers the editor with access to deep static analysis, effectively transforming a standard text editor into a powerful, semantics-aware development environment (more details in Appendix~\ref{details of lsp Protocol}). It powers essential features such as semantic symbol resolution, cross-file reference tracking, and automated refactoring. For instance, when a developer hovers over a function name, LSP provides signature help and documentation; clicking on a symbol triggers a precise jump to its definition or lists all its references across the workspace. These capabilities allow developers to bypass manual text searching, offering real-time, accurate code insights and error detection. By adopting LSP, the software ecosystem gains high scalability, where complex, language-specific logic is implemented once in a Language Server and reused across various editing environments.

\subsubsection{LSP Tools Implementation for Code Agents}

Inspired by the workflow of human developers who rely on intelligent IDEs for efficient debugging and code navigation, we designed and implemented a unified \texttt{lsp\_tool} interface tailored for code agents. Unlike human-facing GUIs, this tool adapts the language server protocol into a function-calling format optimized for LLMs, bridging the gap between raw protocol data and model comprehension.

Our implementation encapsulates a comprehensive suite of features defined in the LSP specification. As presented in Table \ref{tab:lsp_featrues}, we categorize these supported capabilities into four distinct groups: \textit{repo navigation}, \textit{dependency analysis}, \textit{code understanding} and \textit{workspace search}. These features are exposed to the agent via a unified API, allowing it to perform structural exploration of the codebase.

\begin{table}[htbp]
\centering
\small
\renewcommand{\arraystretch}{1.2} 
\resizebox{0.96 \linewidth}{!}{ 
\begin{tabular}{l|l|p{8.7cm}} 
\toprule
\textbf{Category} & \textbf{Feature Name} & \textbf{Description \& Utility for Agents} \\ 
\midrule
\multirow{4}{*}{\textbf{\shortstack[l]{Repo\\Navigation}}} 
 & go to definition & Locate the exact definition of a symbol (function, class, variable). \\
 & go to declaration & Jump to the declaration of a symbol. \\
 & go to type definition & Navigate to the definition of a variable's type. \\
 & go to implementation & Find concrete implementations of an interface or abstract method. \\ 
\midrule
\multirow{3}{*}{\textbf{\shortstack[l]{Dependency\\Analysis}}} 
 & prepare call hierarchy & Initialize the call hierarchy for a selected function. \\
 & incoming calls & Identify all functions that call the target function (Callers). \\
 & outgoing calls & Identify all functions called by the target function (Callees). \\ 
\midrule
\multirow{3}{*}{\textbf{\shortstack[l]{Code\\Understanding}}} 
 & signature help & Provide parameter details and return types for function calls. \\
 & document symbols & Extract a structural outline (tree) of the current file. \\
 & document color & Identify color representations within the code (if applicable). \\ 
\midrule
\multirow{2}{*}{\textbf{\shortstack[l]{Workspace\\Search}}} 
 & workspace symbols & Search for symbols globally across the entire project. \\
 & find references & List all usages of a specific symbol across the workspace. \\
\bottomrule
\end{tabular}
}
\caption{Overview of LSP features integrated into the code agent.}
\label{tab:lsp_featrues}
\end{table}

To facilitate better model interaction, we implemented a parser between the agent and the Language Server, avoiding the direct exposure of complex raw JSON-RPC methods. This parser perform pre-processing and post-processing on the interactions between the model and the Language Server, enabling the agent to interact with the language server losslessly without grappling with the protocol's underlying complexity. The specific detail are provided in appendix~\ref{details of lsp implementation}.

\subsubsection{Workflow Integration}

We integrate the \texttt{lsp\_tool} into the code agent's software engineering task-solving loop, as illustrated in Figure~\ref{fig:lsp-pipline}. In this workflow, the agent operates as a central decision-maker. Upon receiving a task (\eg a GitHub issue), the agent analyzes the repository. Instead of merely adopting brute-force \texttt{find} commands, the agent utilizes \texttt{lsp\_tool} to traverse the codebase structurally—jumping to definitions to understand logic, querying references to assess impact, and analyzing call hierarchies to trace bug propagation. This structured observation is then fed back into the model's context, allowing it to reason about the bug's root cause with high fidelity before generating a patch.


The integration of the \texttt{lsp\_tool} helps transform the agent's operating paradigm from black-box-testing to white-box-analysis. Previously, agents largely relied on a trial-and-error approach—hypothesizing a fix, running scripts, and analyzing error logs—a process prone to redundant, ineffective debugging loops. By empowering the agent to read the code structure directly (\eg via call hierarchies and definition jumps), our approach allows for precise localization of logic defects without the need for constant execution. This shift not only significantly enhances exploration efficiency by reducing the number of invalid interaction turns but also mitigates the risk of hallucinated modifications, ensuring that the agent fully comprehends the code context before attempting a fix.

\subsection{Continual Training of SWE-Master with LSP Tools}

\subsubsection{Verification in Open-Source Foundation Models}

To verify the effectiveness of our proposed toolchain, we conduct experiments on the SWE-bench Verified dataset and select \textbf{pyright} as the underlying language server engine because the dataset is python-only. It is important to note that our system design is modular and language-agnostic; switching to other programming languages (\eg Java or C++) is a ``plug-and-play'' process that requires only replacing the corresponding LSP binary within the Docker container and updating the startup command. In all evaluations involving the \texttt{lsp\_tool}, we enabled the full suite of features listed in Table~\ref{tab:parsed_lsp_tools}, with the exception of \texttt{get\_declaration} and \texttt{get\_implementation}. These two features were omitted solely because the pyright language server does not currently support them for Python's dynamic typing system.

We validate the efficacy of the \texttt{lsp\_tool} on MiniMax-M2.1 and GLM-4.7. As shown in Table~\ref{tab:lsp_comparison}, integrating LSP capabilities yields consistent performance gains across both models. For MiniMax-M2.1, the inclusion of LSP improves the resolution accuracy from 68.4\% to {70.4\%}, while simultaneously reducing the average interaction turns from 82.0 to 77.0. Similarly, GLM-4.7 achieves a higher accuracy of 67.6\% with reduced interaction turns. These results demonstrate that LSP tools effectively lower the cognitive burden on the model: by providing deterministic semantic context instead of noisy keyword search results, the agents can navigate to the bug location more directly, thereby solving tasks more efficiently.

\begin{table}[htbp]
    \small
    \centering
    \begin{tabular}{cccccc}
        \toprule
        \multirow{2}{*}{\textbf{Model}} & \multirow{2}{*}{\textbf{Scaffold}} & \multicolumn{2}{c}{\textbf{w/o LSP}} & \multicolumn{2}{c}{\textbf{w. LSP}} \\
        \cmidrule(lr){3-4} \cmidrule(lr){5-6}
         & & \textbf{Acc.} (\%) & \textbf{Avg. Turns} & \textbf{Acc.} (\%) & \textbf{Avg. Turns} \\
        \midrule
        M2.1 & R2E-Gym & 68.4 & 82.0 & {70.4} & {77.0} \\
        GLM-4.7 & R2E-Gym & 66.2 & 97.3 & {67.6} & {94.4} \\
        \bottomrule
    \end{tabular}
    \caption{Performance comparison of GLM-4.7 and Minimax-M2.1 with and without LSP tools.}
    \label{tab:lsp_comparison}
\end{table}

\subsubsection{Continual Training of SWE-Master}

Motivated by these results on powerful teacher models, we distill the high-level repository navigation skills into SWE-Master, to enable similar IDE-level behaviors. Using the integration method from Section~\ref{sec:traj_gen}, we incorporate LSP tools and leverage GLM-4.6 and Minimax-M2 as teachers to generate trajectories. A rule-based filter combined with an LLM judge ensures only high-quality trajectories demonstrating correct LSP use and problem solving are selected. We then perform SFT starting from the SWE-Master-RL checkpoint, mixing new LSP trajectories with original SFT data to avoid overfitting. Detailed filtering prompts are in Appendix~\ref{p:lsp_filtering_sp}.

As shown in Table~\ref{tab:lsp_comparison_policy_model}, the distilled model achieves significant efficiency gains while maintaining a resolve rate comparable to the strong RL baseline (61.0\% vs. 61.4\%). The integration of LSP capabilities reduces input and output token consumption by {23.7\%} and {16.3\%}, respectively, and shortens the average trajectory by {17.5\%}. This efficiency stems from the agent's shift from verbose, trial-and-error lexical searches to precise semantic queries, achieving equivalent proficiency with substantially lower computational costs. It should be noted, however, that the efficiency gains are not solely attributable to LSP; they also reflect behavioral shifts induced by the continual training of SWE-Master-RL.


\begin{table}[htbp]
    \small
    \centering
    \setlength{\tabcolsep}{5pt} 
    \begin{tabular}{llcccc}
        \toprule
        \textbf{Model} & \textbf{Setting} & \textbf{Input Tok.} & \textbf{Output Tok.} & \textbf{Avg. Turns} & \textbf{Resolve Rate} \\
        & & (k) $\downarrow$ & (k)$\downarrow$  & (k) $\downarrow$ & (\%) $\downarrow$ \\
        \midrule
        \textbf{RL} & w/o LSP. & 5549.6 & 32.5  & 111.2 & 61.4 \\
        \textbf{RL + Cont. SFT} & w. LSP.  & {4232.1} & 27.2 & {91.7} & {61.0} \\
        \bottomrule
    \end{tabular}
    \caption{Comparison of token efficiency and performance metrics of SWE-Master-RL before and after continual training with LSP tools.}
    \label{tab:lsp_comparison_policy_model}
\end{table}

\subsubsection{Case Study on LSP Tool Use}

\begin{figure}[ht]
\centering
    \centering
    \includegraphics[width=1.0\linewidth]{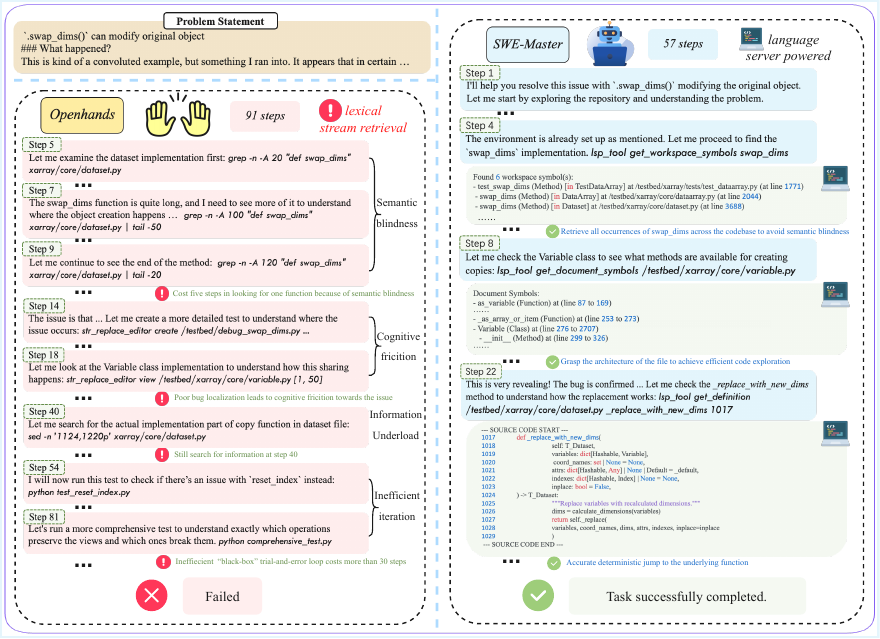}
\caption{Example of a trajectory generated using LSP tools.}
    \label{fig:lsp}
\end{figure}

To explicitly demonstrate how \texttt{lsp\_tool} empower the model to transcend the limitations of lexical search, we present a comparative case study on  \textit{pydata\_xarray-6812} from SWE-bench Verified. The task requires fixing a subtle bug where \texttt{.swap\_dims()} incorrectly modifies the original dataset object in-place due to improper reference handling. Figure~\ref{fig:lsp} visualizes the contrasting trajectories from current frontier framework OpenHands and our enhanced framework integrated with \texttt{lsp\_tool}.

\noindent\textbf{The Struggle of Lexical Retrieval (Baseline).} 
As illustrated on the left side of Figure~\ref{fig:lsp}, the OpenHands agent relies on lexical stream retrieval tools, such as \texttt{grep} and \texttt{find}, for bug localization. In Step 5, it attempts to search for the \texttt{.swap\_dims()} function definition via text matching. However, since \texttt{grep} lacks semantic awareness of function boundaries, the agent is forced to ``scroll'' through the code by repeatedly increasing line counts, wasting 4 steps just to barely piece together a code snippet. This fragmented reading induces \textit{Cognitive Friction}, impeding the agent's ability to comprehend dependence relationships between objects, and to construct a mental model of cross-file dependencies. Lacking a mechanism to jump to definitions, the agent falls into \textit{Information Underload}, still searching for information even in step 40, which means it consumes tokens reading code without grasping the structural relationships. Consequently, the agent falls into a ``black-box'' trial-and-error loop (Steps 54-81), repeatedly running tests without pinpointing the logic error, eventually failing after 91 steps.

\noindent\textbf{The Success of IDE-level Navigation (Ours).} 
In contrast, our enhanced framework (right side of Figure~\ref{fig:lsp}), equipped with the \texttt{lsp\_tool} suite, demonstrates IDE-level intelligent navigation patterns.
\begin{itemize}[leftmargin=1.5em]
    \item \textit{Global Symbol Resolution (Step 4):} Instead of guessing file paths, the agent utilizes \texttt{get\_workspace\_symbols} to bypass massive textual noise. It retrieves all occurrences of \texttt{swap\_dims} across the codebase, thereby constructing a mental model of cross-file dependencies.
    \item \textit{Hierarchical Understanding (Step 8):} By calling \texttt{get\_document\_symbols}, the agent quickly grasps the architecture of the file, specifically the class structure of \texttt{Variable}. It identifies relevant methods like \texttt{to\_index\_variable} without reading the entire file, achieving efficient code exploration.
    \item \textit{Deterministic Trace (Step 22):} The crucial breakthrough occurs when the agent uses \texttt{get\_definition} to trace the \texttt{\_replace\_with\_new\_dims} method. Unlike the baseline which implies relationships from text, LSP provides a deterministic link to the underlying logic. This allows the agent to identify that \texttt{IndexVariable.to\_index\_variable} returns \texttt{self} (a shared reference) rather than a copy, effectively exposing the root cause of the mutability bug.
\end{itemize}

This IDE-level code navigation capability enables SWE-Master to solve the task in just 57 steps—a 37\% reduction in trajectory length—proving that the distilled model has successfully learned to read code structure rather than just search text.
\section{Further Analysis}

\subsection{Data Scaling for SFT}\label{sec:ana:sft_data_scaling}

\begin{figure}[ht]
\centering
    \centering
    \includegraphics[width=0.9\linewidth]{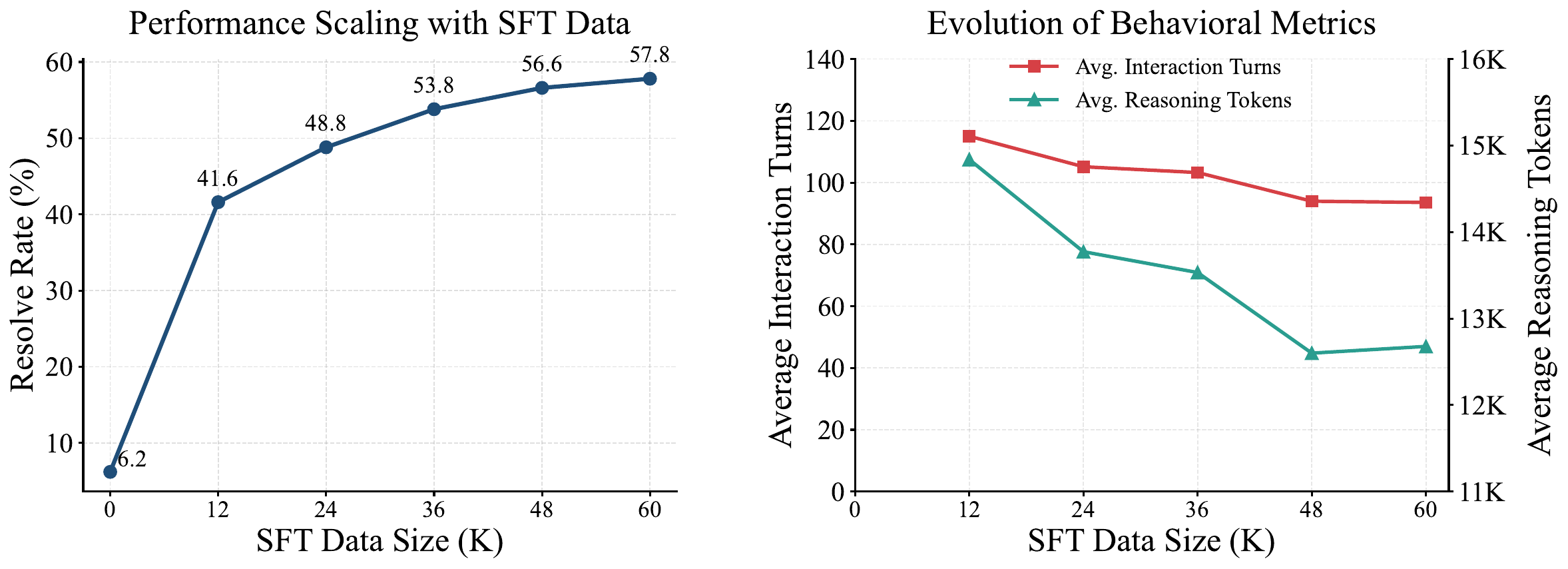}
\caption{Impact of SFT data scaling on performance and reasoning efficiency. }
    \label{fig:sft_data_scaling}
\end{figure}

To investigate the data scaling laws within the multi-turn SFT phase, we evaluate the model's performance and behavioral evolution as the training corpus expands from 0 to 60K samples. As illustrated in Figure~\ref{fig:sft_data_scaling} (Left), the resolve rate exhibits a robust logarithmic growth, surging from a baseline of 6.2\% to 57.8\%; however, the marginal utility of additional data begins to plateau beyond 48K, suggesting a saturation point for supervised behavior cloning. Crucially, this performance improvement is accompanied by a marked increase in inference efficiency, as shown in Figure~\ref{fig:sft_data_scaling} (Right). Both the average interaction turns and the thinking token consumption demonstrate a consistent downward trend, decreasing from approximately 115 to 94 turns and 14.8k to 12.7k tokens, respectively. This inverse correlation implies that as the model absorbs more expert demonstrations, it internalizes more efficient problem-solving heuristics, thereby reducing redundant exploration and ineffective reasoning loops while achieving higher accuracy with fewer computational resources.

\subsection{Data Filtering for SFT}\label{sec:ana:data_filter}

To assess the impact of data quality control, we examine the performance differential between models trained with and without difficulty-based filtering as shown in Section~\ref{sec:data_filting}.  Specifically, \textit{w. difficulty-based filtering} excludes problems with invariant outcomes in which all corresponding trajectories either consistently pass or consistently fail, whereas \textit{w/o difficulty-based filtering} involves random sampling directly from the pool of all correct trajectories.
As presented in Table~\ref{tab:bon_ablation}, applying the filtering mechanism yields a distinct improvement in the resolve rate, suggesting that theverified, high-quality trajectories is essential for model training. Notably, this performance gain is achieved with negligible variations in average interaction turns and thinking token consumption. It refines the correctness of the agent's reasoning, enabling it to resolve issues more reliably without altering the fundamental behavioral complexity or computational cost.

\begin{table}[ht]
    \centering
    \small
    \caption{Ablation study on the effectiveness of difficulty-based filtering.}
    \label{tab:bon_ablation}
    \setlength{\tabcolsep}{6pt} 
    \resizebox{0.95\linewidth}{!}{
    \begin{tabular}{c c c c}
        \toprule
        \textbf{Method} & \textbf{Resolve Rate (\%)} & \textbf{Avg. Turns} & \textbf{Avg. Think Tokens} \\
        \midrule
        w/o difficulty-based Filtering & 54.2 & 94.57 & 12,850 \\
        w. difficulty-based Filtering  & {57.8} & {93.58} & {12,678} \\
        \bottomrule
    \end{tabular}
    }
\end{table}

\subsection{Loss Masking and Reward Design in RL}\label{sec:ana:reward_design}

DeepSWE~\citep{deepswe2025} suggests that setting the reward to zero and masking the loss for trajectories truncated by environmental constraints (\eg maximum context, timeouts, or interaction turn limits) enhances stability during the RL phase. However, contrary to these findings, our experiments indicate that applying this masking strategy to SWE-Master precipitates a continuous degradation in reward and eventual training collapse, rather than fostering efficient reasoning. We attribute this discrepancy to fundamental differences in policy model initialization and data complexity: whereas DeepSWE trains from scratch on the relatively simpler R2E-Gym dataset (as detailed in Section~\ref{sec:traj_gen} and Figure~\ref{fig:data_rollout_acc_distribution}), SWE-Master undergoes cold-start stage through sufficient SFT on long-horizon heterogeneous data, resulting in a distinct optimization landscape. 

\begin{figure}[ht]
\centering
    \centering
    \includegraphics[width=0.9\linewidth]{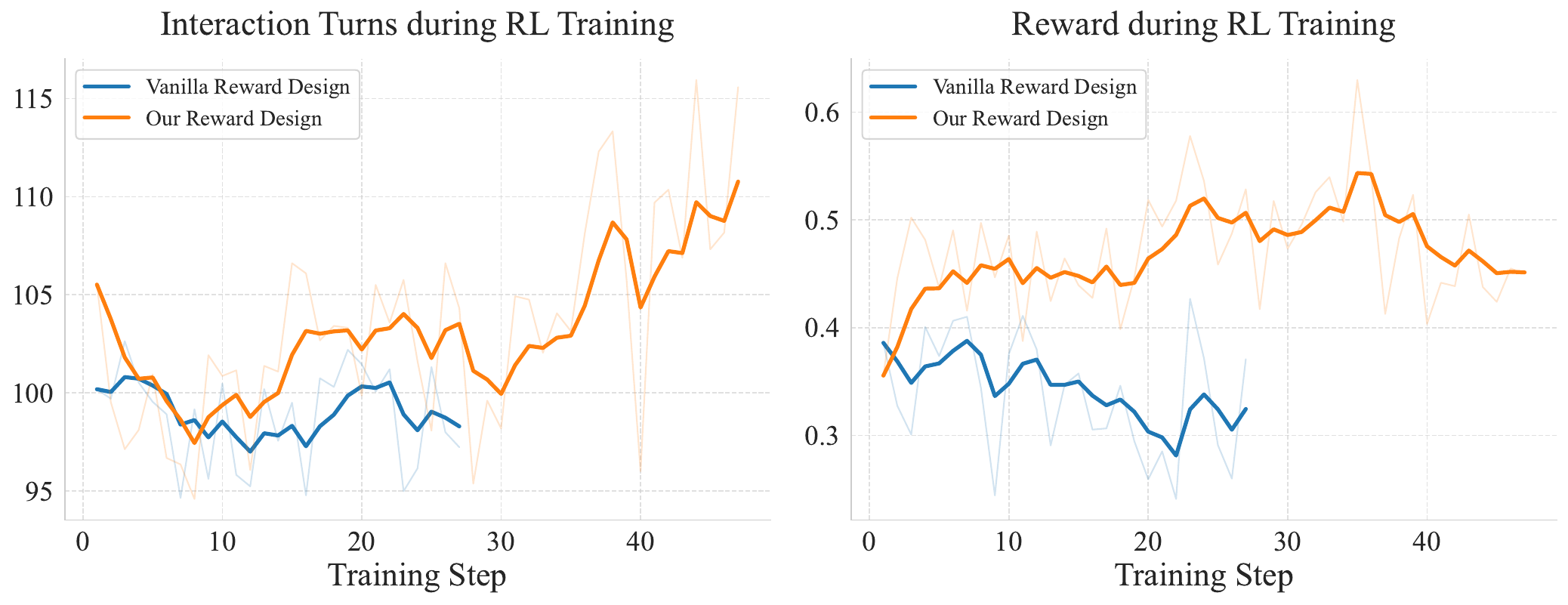}
\caption{ Comparison of interaction turns, reward, and entropy between DeepSWE and our proposed loss masking and reward design strategies based on the SWE-Master-SFT model.}
    \label{fig:reward_design_compare}
\end{figure}

This divergence is empirically illustrated in Figure~\ref{fig:reward_design_compare}, which compares the training dynamics of the DeepSWE's masking strategy (blue line) against our reward shaping mechanism (orange line). While the masking approach leads to instability, our method ensures stable reward growth and facilitates deeper interaction, as evidenced by the increasing trend in trajectory length. Although this encouragement of longer reasoning chains imposes a slight penalty on training throughput due to increased computational overhead, it proves indispensable for achieving robust convergence on complex software engineering tasks.

\subsection{Git Hacking}\label{sec:ana:git_hacking}

\begin{figure}[ht]
\centering
    \centering
    \includegraphics[width=0.8\linewidth]{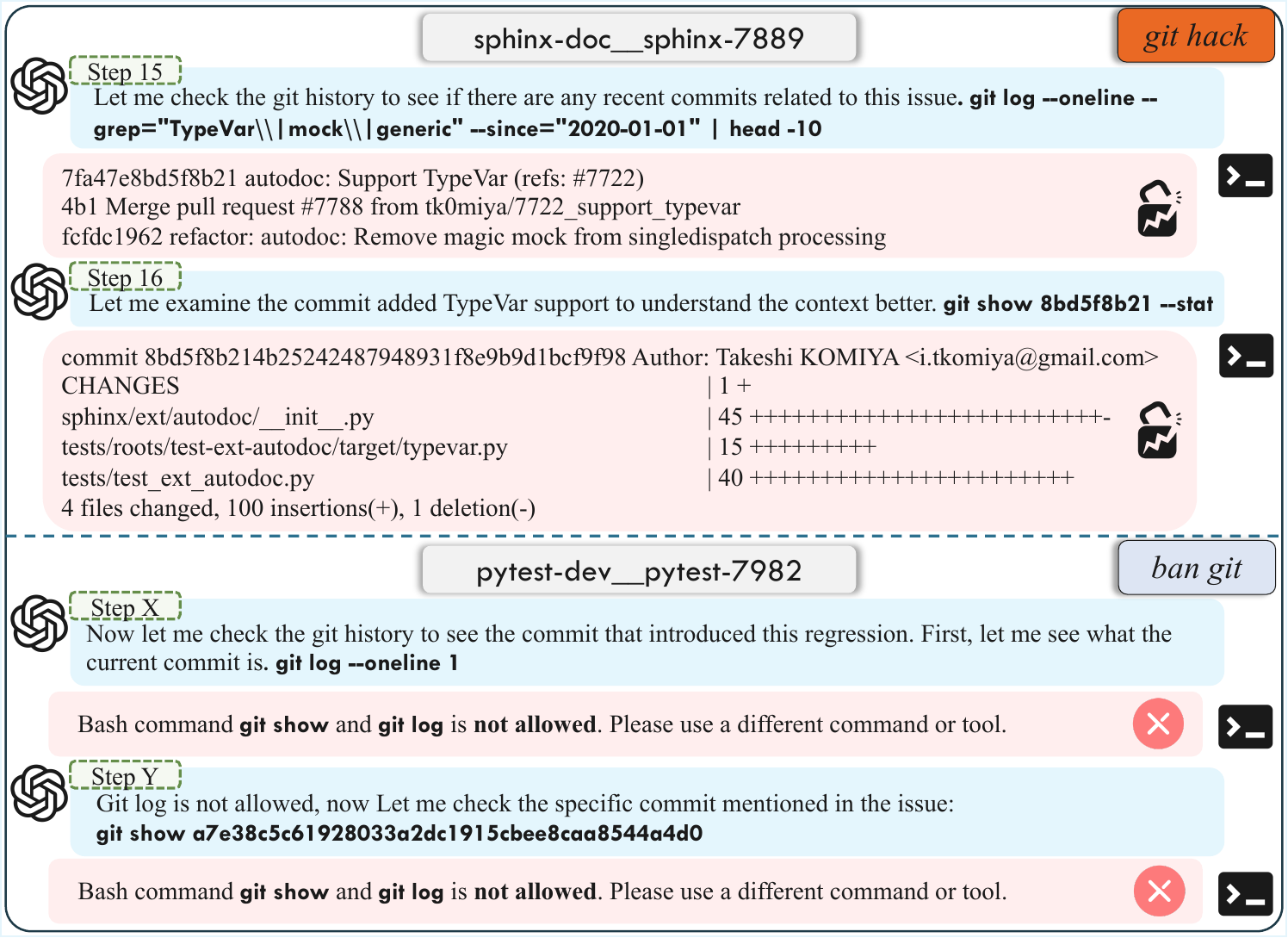}
    \caption{Comparison between the model's autonomous \textit{git hack} attempts (top), such as \texttt{git log} and \texttt{git show}, and the corresponding environmental intervention (bottom).}
    \label{fig:git_hack_compare}
\end{figure}

Consistent with observations in MiMo-v2-Flash~\citep{xiao2026mimo}, we found that in SWE scenarios, the model autonomously attempts to exploit commands such as \texttt{git show} and \texttt{git log} to retrieve golden patches directly from the network, as shown in Figure~\ref{fig:git_hack_compare}. To address this, as detailed in Section~\ref{sec:framework_and_env}, the environment is configured to intercept such attempts and return a prohibition warning whenever the model seeks to extract solutions via unauthorized network access. This constraint was strictly enforced during both the SFT and RL stages.

We conduct an ablation study on SWE-Master to evaluate the impact of this mandatory anti-hacking mechanism, with results presented in Table~\ref{tab:git_ablation}. Interestingly, removing the restriction on Git commands results in a slight degradation in performance. We attribute this phenomenon to the model's lack of exposure to such exploitation patterns during the training phase; consequently, the trained model lacks proficiency in utilizing these hacking tools effectively, rendering these attempts less successful than standard problem-solving approaches. These findings further validate the effectiveness of our tool-level strategy in preventing data leakage and ensuring robust evaluation.

\begin{table}[htbp]
    \centering
    \small
    \caption{Performance comparison with and without Git tools, along with a focused analysis of identified \textit{git hacking} behaviors.}
    \label{tab:git_ablation}
    \setlength{\tabcolsep}{8pt} 
    \resizebox{0.95\linewidth}{!}{
    \begin{tabular}{c c | c c c}
        \toprule
        \multirow{2}{*}{\textbf{Method}} & \textbf{w/o Git Tool} & \textbf{w/ Git Tool} & \multicolumn{2}{c}{\textbf{Hacking Analysis}} \\
        & \textbf{Resolve Rate (\%)} & \textbf{Resolve Rate (\%)} & \textbf{\# Samples} & \textbf{Acc. (\%)} \\
        \midrule
        SFT & 57.8 & 57.0 & 51 & 56.8 \\
        RL  & 61.4 & 58.4 & 28 & 57.1 \\
        \bottomrule
    \end{tabular}
    }
\end{table}

\subsection{Summary-Based Context Manager for SWE Tasks}
\label{subsec:context_manager}

The SWE tasks require LLMs to continuously interpret environment feedback, execute actions, and revise strategies over numerous interaction rounds~\citep{wang2026memgovernenhancingcodeagents}. This imposes significant demands on context management. Currently, leading code agents and frameworks~\citep{liu2025deepseek, zeng2025glm}, predominantly adopt an \textit{Append-Only} strategy. In this approach, all historical interaction information is sequentially concatenated to the message history without compression.

Conversely, in web search scenarios, approaches like the \textit{Condenser}, described as Discard-All~\citep{liu2025deepseek}, have achieved frontier performance on BrowseComp~\citep{wei2025browsecomp} benchmarks. This method automatically discards all tool response from previous turns once a specific window size is exceeded. While effective for search, we argue that this approach is detrimental in coding scenarios. Because coding tasks are heavily state-dependent, discarding observations (\eg file contents read in previous turns, error logs) strips the model of critical environmental context, forcing it to redundantly re-read files or hallucinate code structures.

\begin{figure}[htbp]
\centering
    \centering
    \includegraphics[width=0.95\linewidth]{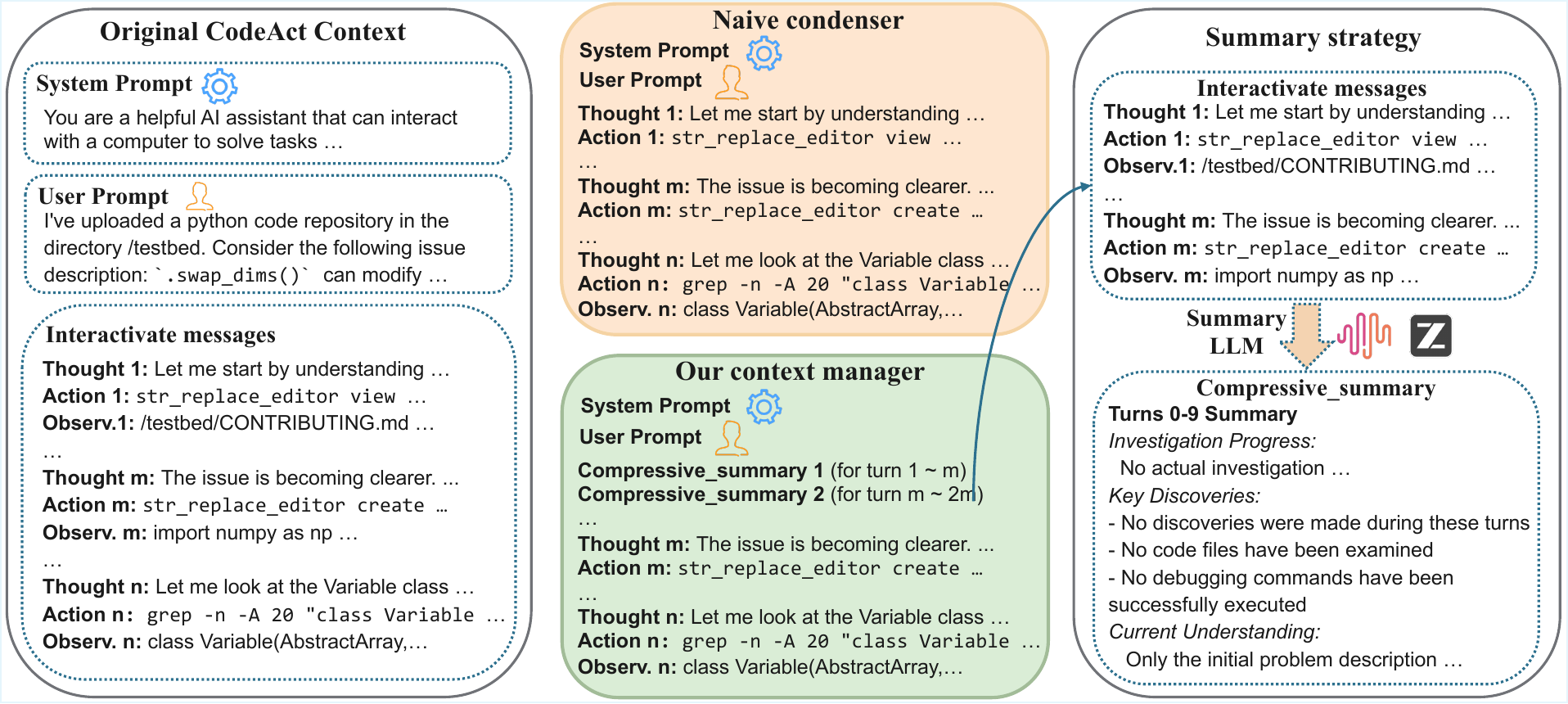}
\caption{Comparison of different context management strategies.}
    \label{fig:context_manager}
\end{figure}

To bridge this gap, we propose a summary-based context manager specifically optimized for SWE scenarios inspired by summary methods~\citep{huang2025manusearch} adopted in search scenarios. As shown in Figure~\ref{fig:context_manager}, instead of discarding history, we compress past interactions into natural language summaries while maintaining a high-fidelity sliding window for recent turns. 
Formally, let $m$ be the summarization interval (\ie a summary is triggered every $m$ turns) and $k$ be the minimum size of the recent context window reserved for raw interactions. At any given turn $t$, the number of completed summary blocks is $L = \lfloor (t - k) / m \rfloor$. The context $\tau_t$ is constructed as:
\begin{equation}
    \tau_t = (I, S_1, S_2, \dots, S_L, \underbrace{(a_{t-w_t}, o_{t-w_t}), \dots, (a_{t-1}, o_{t-1})}_{\text{Recent Sliding Window } w_t})
\end{equation}
where $w_t= t - L \cdot m$ represents the raw context window, dynamically varying between $k$ and $k + m - 1$. The summarization process is triggered only when the window size $w_t$ reaches $k+m$. At that point, the oldest $m$ turns in the window are condensed into a new summary $S_{L+1}$ by a summarizer model $\pi_{summary}$:
\begin{equation}
    S_{L+1} = \pi_\text{summary}\left( \{(a_j, o_j)\}_{j=L \cdot m}^{(L+1) \cdot m - 1} \right)
\end{equation}

This hybrid structure ensures that the model retains long-term memory of high-level logic while possessing precise, token-level access to the immediate context required for debugging and editing. 

We integrate this strategy into our framework and evaluate it on the SWE-bench Verified dataset using MiniMax-M2.1 and GLM-4.7, with each model serving as its own summary model. As presented in Table~\ref{tab:memory_performance}, our context manager significantly enhances efficiency without compromising capabilities.
For the M2.1 model, enabling the context manager yields a 42\% saving in total input tokens (from 2.8M to 1.6M) and reduces the average peak token usage per trajectory by nearly 50\% (from 55.4k to 28.5k).This optimization effectively enables the model to transcend its inherent context length limitations. Crucially, this aggressive compression does not degrade performance, likely because the cleaner context helps the model focus on relevant information. A similar trend is observed with GLM-4.7, where token consumption drops by 36\% while maintaining a competitive resolve rate. The increase in average turns suggests that the reduced context overhead allows the agent to explore deeper and attempt more correction steps before hitting context limits. 
However, when extending this mechanism to SWE-Master, we observed no measurable improvements in either efficiency or efficacy.  We attribute this to the model's smaller parameter scale, which limits its ability to effectively utilize the managed context. 

\begin{table}[htbp]
    \small
    \centering
    \setlength{\tabcolsep}{5pt} 
    \begin{tabular}{ccccccc}
        \toprule
        \textbf{Model} & \textbf{Setting} & \textbf{Input Tok} & \textbf{Output Tok} & \textbf{Avg Peak Tok} & \textbf{Avg Turns} & \textbf{Resolve Rate} \\
        & & (k) $\downarrow$ & (k)$\uparrow$  & (k) $\downarrow$ &$\uparrow$ & (\%)  \\
        \midrule
        \multirow{2}{*}{\textbf{M2.1}} 
        & w/o Mem. & 2821.9 & 18.7 & 55.4 & 82.0 & 68.4 \\
        & w. Mem.  & {1636.6} & 24.9 & {28.5} & {92.9} & {69.8} \\
        \midrule
        \multirow{2}{*}{\textbf{GLM-4.7}} 
        & w/o Mem. & 3540.4 & 20.9 & 63.3 & 97.3 & {66.2} \\
        & w. Mem.  & {2259.8} & 28.5 & {32.5} & {120.0} & 65.8 \\
        \bottomrule
    \end{tabular}
    \caption{Comparison of token efficiency and performance metrics based on Minimax-M2.1 and GLM-4.7. {Input Tok} and {Output Tok} refer to the average all input and output all tokens across all evaluation trajectories, respectively.  {Avg Peak Tok} refers to the average maximum single-turn context length per trajectory.
}
    \label{tab:memory_performance}
\end{table}


\section{Related Work}

\paratitle{Datasets for Software Engineering Tasks.}
The evaluation paradigm for code models has witnessed a significant shift, transitioning from standalone algorithmic code generation~\citep{chen2021evaluating, jain2024livecodebench} to complex, repository-level software engineering tasks. SWE-bench~\citep{jimenez2024swebench} represents a pioneering milestone in this domain, serving as the first comprehensive benchmark to assess the capability of LLM-based code agents in resolving real-world GitHub issues. Building upon this foundation, {SWE-bench Verified}~\citep{sweb-verified} is subsequently introduced, providing a refined subset that rigorously ensures issue solvability.
In parallel, to enhance the proficiency of code agents in repository-level issue resolution, recent research has focused on the construction of high-quality training datasets and execution environments. Works such as {SWE-Gym}~\citep{swe-gym}, {R2E-Gym}~\citep{jain2025r2e}, {SWE-smith}~\citep{yang2025swesmith}, and {SWE-rebench}~\citep{badertdinov2025swerebench} have established scalable pipelines for harvesting, constructing, and filtering data from GitHub, thereby creating robust environments for code agent training. Furthermore, the landscape of agent evaluation is rapidly expanding to cover broader dimensions. Recent initiatives\citep{xu2025swecompassunifiedevaluationagentic}, including {Multi-SWE-bench}~\citep{zan2025multiswebench} and {SWE-bench Pro}\citep{Deng2025SWEBenchPC} extend the scope to diverse programming languages and introduce heightened difficulty issues, while SWE-bench Multimodal~\cite{yang2025swebenchmultimodal} extends the issue-solving paradigm to the multimodal domain. Furthermore, Terminal-Bench~\cite{merrill2026terminalbenchbenchmarkingagentshard} introduces diverse task types, facilitating a more comprehensive and multi-faceted evaluation of code agents' capabilities.

\paratitle{Software Engineering LLMs and Agents.}
To enhance the capabilities of autonomous code agents, researchers have pursued several distinct paradigms. Early efforts primarily focused on the development of agentic frameworks. Representative systems such as SWE-agent~\citep{yang2024swe} and OpenHands~\citep{wang2025openhands} utilize powerful foundational models~\citep{liu2025deepseek,wong2025confuciuscodeagentscalable} within a real-world environment, demonstrating strong performance in issue-solving tasks.
Meanwhile, recent studies have shifted toward enhancing the underlying models specifically for software engineering tasks. For instance, daVinci-Dev~\citep{zeng2026davinci} employs data synthesis and mid-training to instill agentic reasoning into base models. Similarly, SWE-Mirror~\citep{wang2025swe} and SWE-Lego~\citep{tao2026swe} leverage agentic supervised fine-tuning~\citep{sun2025simpledeepsearcherdeepinformationseeking} by generating and sampling high-quality trajectories from proprietary LLMs within grounded execution environments. To further bridge the gap between static training and dynamic interaction, DeepSWE~\citep{deepswe2025} and SkyRL~\citep{cao2025skyrl} incorporate agentic reinforcement learning~\citep{song2025r1,golubev2025training}. These approaches optimize agent policies by allowing them to learn from trial-and-error feedback within sandboxed container environments.
In contrast to the end-to-end agentic loop, a parallel line of research explores agentless pipelines~\citep{xia2024agentless, yang2025kimi}. Instead of maintaining a continuous autonomous state, agentless pipelines decompose the issue-solving task into three discrete, predefined stages: fault localization, code repair, and patch verification~\citep{xie2025swe, SWESwiss2025}, and each stage is optimized independently.

\section{Conclusion}

In this work, we present {SWE-Master}, an open-source software engineering agents capable of resolving complex, repository-level issues. By building a robust infrastructure with decoupled Docker execution environments and adopting a carefully designed data curation pipeline—including agent-based trajectory rollout, format-based filtering, and difficulty-based selection—we construct a high-quality dataset that substantially improves training effectiveness for both SFT and RL.
On top of this foundation, we propose a reinforcement learning framework tailored to long-horizon software engineering tasks. In addition, we reduce the semantic gap in code navigation by introducing a novel, language-agnostic \texttt{lsp\_tool}, which provides IDE-level structural information to the agent.
The evaluations on SWE-bench Verified show that SWE-Master achieves advanced performance among open-source code agents. These results demonstrate the effectiveness of our end-to-end approach, spanning environment design, data engineering, and advanced reinforcement learning techniques, in advancing autonomous software engineering agents.


\bibliographystyle{unsrt}
\bibliography{ref.bib}

\newpage
\appendix

\section{Detailed Implementation of LSP Tools}
\label{app:lsp}
This section complements Section~\ref{sec:ide_injection} with a more fine-grained details of the implementation of LSP Tool, along with a fuller characterization of the tool call format.

\subsection{Language Server Protocol}
\label{details of lsp Protocol}

\begin{figure}[ht]
\centering
    \centering
    \includegraphics[width=0.6 \linewidth]{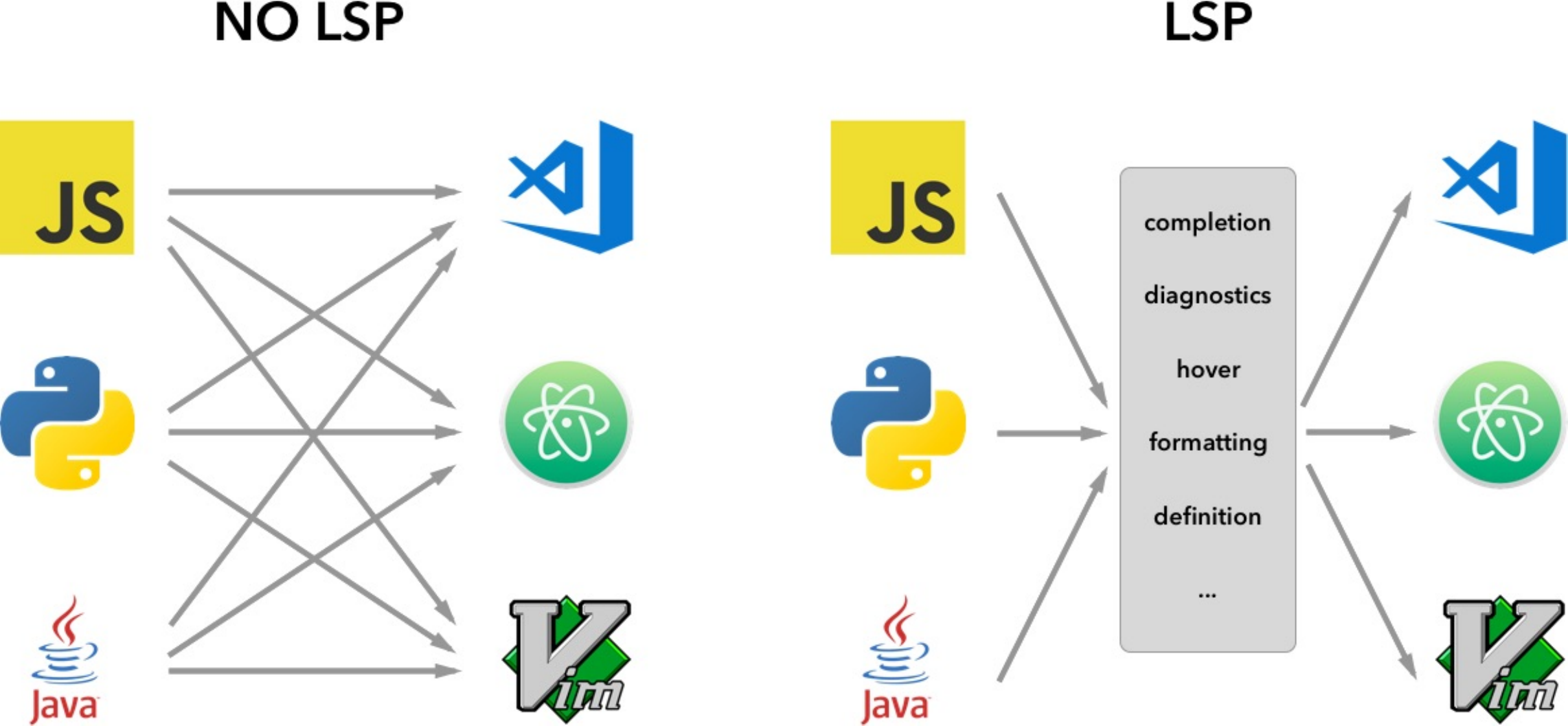}
\caption{The Language Server Protocol for use between integrated development environments (IDEs) and servers that provide ``language intelligence tools''.}
    \label{fig:lsp-language-editors}
\end{figure}

As shown in Figure~\ref{fig:lsp-language-editors}, the core architectural contribution of the Language Server Protocol is the mitigation of the $M$-to-$N$ point-to-point integration bottleneck. Historically, providing support for $N$ programming languages across $M$ different editors required $M \times N$ unique implementations. By introducing a standardized intermediate protocol, LSP transforms this requirement into a more scalable $M + N$ problem, where a single language server can serve any compliant client, significantly lowering the barrier for language support in diverse development environments.

There are three components of LSP, including Language Server (LS), Language Client (LC) and the protocol~\citep{understand-lsp}. Language Server is a process that provides language smarts by communicating with development tools. Language Client is a code editor/development tool or an extension that can communicate with a particular Language Server. To deliver intelligent language features, the Language Server functions essentially as a persistent compiler frontend.  Upon receiving source code updates from the client, the server performs lexical and syntactic analysis to construct an Abstract Syntax Tree (AST), a hierarchical representation of the code's structure. Traversing this AST, the server builds and maintains a Symbol Table, which resolves identifier scopes, types, and bindings. To support workspace-wide operations, the server further aggregates these symbols into a Symbol Search Engine (or global index). It is through the synthesis of the AST, Symbol Table, and Search Engine that the server translates raw text positions from JSON-RPC requests into semantic responses. The protocol is analogous to HTTP, which consists of a header part and a content part. The header part consists of two header fields named content length and content type. Content length is a number indicating the length of the content part in bytes. And the content type indicates the mime type of the content part which defaults to \texttt{application/vscode-jsonrpc; charset=utf-8} which is used to encode the content part. More details in~\citep{lsp}.

The lifecycle of an LSP session begins when the Language Client spawns the Language Server process and establishes a communication channel, typically utilizing standard I/O or TCP sockets. The interaction formally commences with an initialization handshake, wherein the client transmits an \texttt{initialize} request detailing its specific capabilities and workspace context. In response, the server returns an \texttt{InitializeResult} to negotiate and declare its supported language features, such as incremental text synchronization or completion providers. Once the session is established, it enters an operational loop driven by user actions: the client sends JSON-RPC \textit{notifications} (\eg \texttt{textDocument/didChange}) to update the server's internal state without expecting a reply, and \textit{requests} (\eg \texttt{textDocument/definition}) to query the server's static analysis engine. The server processes these queries and returns responses that the client renders in the UI. Finally, the session concludes with a \texttt{shutdown} request to ensure graceful resource cleanup, followed by an \texttt{exit} notification to terminate the server process.

It is precisely through this structured workflow that the Language Server Protocol establishes a unified mechanism for interaction and code intelligence delivery between the server and the client. By enabling this seamless exchange of semantic data, LSP facilitates the significant leap from a text editor to the Integrated Development Environment.

\subsection{LSP Features Integrated into the Code Agent}
\label{details of lsp implementation}
In the \textbf{pre-processing} phase, we expose a simplified, intuitive function-calling interface to the model and translate these calls into standard JSON-RPC requests. Notably, we aggregated the three distinct call hierarchy methods defined in the LSP specification into a single, unified feature—\texttt{get\_call\_hierarchy}—to streamline the analysis workflow, thereby reducing token consumption and cognitive load.

In the \textbf{post-processing} phase, we parse the server's raw JSON responses into a format readable by the model. To maximize the validity and density of the returned information, we selectively augment certain outputs with necessary context (\eg fetching the actual source code content rather than just location coordinates) while filtering out redundant noise. Table~\ref{tab:parsed_lsp_tools} details the specific input parameters and output results for our processed tools, while the Appendix provides a comparison with the standard raw JSON-RPC format.

\begin{table}[htbp]
\centering
\footnotesize
\begin{tabularx}{0.98 \textwidth}{p{2.2cm} l l l}
\toprule
\textbf{Category} & \textbf{Tool Call Name} & \textbf{Input Paras.} & \textbf{Output Results} \\
\midrule

\multirow{4}{=}{Repo Navigation}
& \texttt{get\_definition} & \texttt{fp, l, s} & The symbol's source code. \\
& \texttt{get\_declaration} & \texttt{fp, l, s} & The symbol's source code. \\
& \texttt{get\_type\_definition} & \texttt{fp, l, s} & The symbol's source code. \\
& \texttt{get\_implementation} & \texttt{fp, l, s} & The symbol's source code. \\
\midrule

Dependency Ana.
& \texttt{get\_call\_hierarchy} & \texttt{fp, l, s} & The function's call hierarchy. \\
\midrule

\multirow{3}{=}{Code Understanding}
& \texttt{get\_hover} & \texttt{fp, l, s} & The function signature. \\
& \texttt{get\_document\_symbols} & \texttt{fp} & The document outline. \\
& \texttt{get\_document\_highlights} & \texttt{fp, l, s} & Highlight usages of the symbol. \\
\midrule

\multirow{2}{=}{Workspace Search}
& \texttt{get\_workspace\_symbols} & \texttt{query} & All symbol information globally. \\
& \texttt{get\_references} & \texttt{fp, l, s} & All symbol references globally. \\
\bottomrule
\end{tabularx}
\caption{Overview of LSP Features Integrated into the Code Agent. \texttt{fp} is an abbreviation of  \texttt{file\_path}. \texttt{l} is an abbreviation of  \texttt{line}, the correct line number of the selected symbol in the code file. \texttt{s} is an abbreviation of  \texttt{symbol}, the name of a code entity, including identifiers for Classes, Functions, Methods, Variables, Fields, and Modules. }
\label{tab:parsed_lsp_tools}

\end{table}

\subsection{Tool Call Format of LSP Tool}
We show eight of the ten features used in the evaluation of SWE-bench Verified,  excluding \texttt{get\_declaration} and \texttt{get\_implementation}. These two features are not shown because the Pyright language server does not currently support them due to Python's dynamic typing system, although they are available for languages like C++ or Java and other programming languages.

\definecolor{bg_gray}{RGB}{245,245,245}
\definecolor{frame_blue}{HTML}{A9D18E}  
\definecolor{xml_tag}{RGB}{0,0,128}
\definecolor{xml_attr}{RGB}{128,0,0}

\definecolor{gray_1}{HTML}{B7B7B7}
\definecolor{gray_2}{HTML}{F0F0F0} 
\definecolor{frame_blue}{HTML}{A9D18E}

\lstset{
    basicstyle=\ttfamily\small,
    breaklines=true, 
    columns=fullflexible,
    keepspaces=true,
    frame=none,
    backgroundcolor=\color{gray_2},
    aboveskip=0pt,
    belowskip=0pt
}


\newtcolorbox[auto counter]{LspExample}[1]{
    enhanced,
    breakable, 
    colback=gray_2, 
    colframe=gray_1, 
    coltitle=white, 
    fonttitle=\bfseries,
    title={Example \thetcbcounter: #1}, 
    arc=2pt, 
    boxrule=1pt,
    left=2mm, right=2mm, top=2mm, bottom=2mm,
    before upper={\textbf{\textsf{Input:}}\par\smallskip},
    segmentation style={solid, color=gray_1, line width=1pt},
}

\newcommand{\lspseparator}{
    \tcbline
    \textbf{\textsf{Output:}}\par\smallskip
}

\begin{LspExample}{get\_workspace\_symbols}
\begin{lstlisting}
<function=lsp_tool>
<parameter=command>get_workspace_symbols</parameter>
<parameter=query>swap_dims</parameter>
</function>
\end{lstlisting}

\lspseparator 

\begin{lstlisting}
Execution output of [lsp_tool]:
[status_code]:
success
[Result]:
Found 6 workspace symbol(s):
swap_dims (Method) [in DataArray] at /testbed/xarray/core/dataarray.py  (line 2044)
swap_dims (Method) [in Dataset] at /testbed/xarray/core/dataset.py  (line 3688)
test_append_with_append_dim_not_set_raises (Method) [in ZarrBase] at /testbed/xarray/tests/test_backends.py  (line 2095)
test_no_warning_from_open_emptydim_with_chunks (Method) [in ZarrBase] at /testbed/xarray/tests/test_backends.py  (line 2231)
test_swap_dims (Method) [in TestDataArray] at /testbed/xarray/tests/test_dataarray.py  (line 1771)
test_swap_dims (Method) [in TestDataset] at /testbed/xarray/tests/test_dataset.py  (line 2958)
\end{lstlisting}
\end{LspExample}

\vspace{1em} 

\begin{LspExample}{get\_definition}
\begin{lstlisting}
<function=lsp_tool>
<parameter=command>get_definition</parameter>
<parameter=file_path>/testbed/xarray/core/dataset.py</parameter>
<parameter=symbol>_replace_with_new_dims</parameter>
<parameter=line>1017</parameter>
</function>
\end{lstlisting}

\lspseparator 

\begin{lstlisting}
Execution output of [lsp_tool]:
[status_code]:
success
[Result]:
Found definition at:
/testbed/xarray/core/dataset.py  (start at line 1017 char 9, end at line 1017 char 31)
--- SOURCE CODE START ---
1017            def _replace_with_new_dims(
1018                self: T_Dataset,
1019                variables: dict[Hashable, Variable],
1020                coord_names: set | None = None,
1021                attrs: dict[Hashable, Any] | None | Default = _default,
1022                indexes: dict[Hashable, Index] | None = None,
1023                inplace: bool = False,
1024            ) -> T_Dataset:
1025                """Replace variables with recalculated dimensions."""
1026                dims = calculate_dimensions(variables)
1027                return self._replace(
1028                    variables, coord_names, dims, attrs, indexes, inplace=inplace
1029                )
--- SOURCE CODE END ---
\end{lstlisting}
\end{LspExample}

\begin{LspExample}{get\_type\_definition}
\begin{lstlisting}
<function=lsp_tool>
<parameter=command>get_type_definition</parameter>
<parameter=file_path>/R2E-Gym/src/r2egym/agenthub/runtime/docker.py</parameter>
<parameter=symbol>logger</parameter>
<parameter=line>342</parameter>
</function>
\end{lstlisting}

\lspseparator 

\begin{lstlisting}
[status_code]: 
 success 
[Result]: 
Found definition at:
- /home/usr/anaconda3/lib/python3.12/site-packages/pyright/dist/dist/typeshed-fallback/stdlib/logging/__init__.pyi  (start at line 114 char 7, end at line 114 char 13)
- /home/usr/anaconda3/lib/python3.12/logging/__init__.py  (start at line 1482 char 7, end at line 1482 char 13)

--- SOURCE CODE START ---
   114  class Logger(Filterer):
   115      name: str  # undocumented
   116      level: int  # undocumented
   117      parent: Logger | None  # undocumented
   118      propagate: bool
   119      handlers: list[Handler]  # undocumented
   120      disabled: bool  # undocumented
   121      root: ClassVar[RootLogger]  # undocumented
   122      manager: Manager  # undocumented
   123      def __init__(self, name: str, level: _Level = 0) -> None: ...
   124      def setLevel(self, level: _Level) -> None: ...
   125      def isEnabledFor(self, level: int) -> bool: ...
   126      def getEffectiveLevel(self) -> int: ...
   127      def getChild(self, suffix: str) -> Self: ...  # see python/typing#980
   128      if sys.version_info >= (3, 12):
   129          def getChildren(self) -> set[Logger]: ...
   130
    ......
   220      def makeRecord(
   221          self,
   222          name: str,
   223          level: int,
   224          fn: str,
   225          lno: int,
   226          msg: object,
   227          args: _ArgsType,
   228          exc_info: _SysExcInfoType | None,
   229          func: str | None = None,
   230          extra: Mapping[str, object] | None = None,
   231          sinfo: str | None = None,
   232      ) -> LogRecord: ...
   233      def hasHandlers(self) -> bool: ...
   234      def callHandlers(self, record: LogRecord) -> None: ...  # undocumented

--- SOURCE CODE END ---
\end{lstlisting}
\end{LspExample}

\begin{LspExample}{get\_call\_hierarchy}
\begin{lstlisting}
<function=lsp_tool>
<parameter=command>get_call_hierarchy</parameter>
<parameter=file_path>/testbed/sympy/combinatorics/perm_groups.py</parameter>
<parameter=symbol>minimal_blocks</parameter>
<parameter=line>4354</parameter>
</function>
\end{lstlisting}

\lspseparator 

\begin{lstlisting}
[status_code]:
 success
[Result]:
Call Hierarchy Analysis for: minimal_blocks
Location:  file:///testbed/sympy/combinatorics/perm_groups.py

=== Incoming Calls (Who calls this?) ===
- Caller: sylow_subgroup (Function)
  Location: /testbed/sympy/combinatorics/perm_groups.py line 4258
- Caller: test_minimal_blocks (Function)
  Location: /testbed/sympy/combinatorics/tests/test_perm_groups.py line 514

=== Outgoing Calls (Who does this call?) ===
- Callee: len (Function)
  Location: /opt/miniconda3/envs/testbed/lib/python3.9/site-packages/pyright/dist/dist/typeshed-fallback/stdlib/builtins.pyi line 1572
- Callee: range (Class)
  Location: /opt/miniconda3/envs/testbed/lib/python3.9/site-packages/pyright/dist/dist/typeshed-fallback/stdlib/builtins.pyi line 1334
- Callee: tuple (Class)
  Location: /opt/miniconda3/envs/testbed/lib/python3.9/site-packages/pyright/dist/dist/typeshed-fallback/stdlib/builtins.pyi line 1043
- Callee: is_transitive (Method)
  Location: /testbed/sympy/combinatorics/perm_groups.py line 2334
- Callee: schreier_vector (Method)
  Location: /testbed/sympy/combinatorics/perm_groups.py line 3702
- Callee: append (Method)
  Location: /opt/miniconda3/envs/testbed/lib/python3.9/site-packages/pyright/dist/dist/typeshed-fallback/stdlib/builtins.pyi line 1126
- Callee: random_stab (Method)
  Location: /testbed/sympy/combinatorics/perm_groups.py line 3311
- Callee: PermutationGroup (Class)
  Location: /testbed/sympy/combinatorics/perm_groups.py line 24
- Callee: stabilizer (Method)
  Location: /testbed/sympy/combinatorics/perm_groups.py line 3749
- Callee: orbits (Method)
  Location: /testbed/sympy/combinatorics/perm_groups.py line 2850
- Callee: minimal_block (Method)
  Location: /testbed/sympy/combinatorics/perm_groups.py line 2481
- Callee: _number_blocks (Function)
  Location: /testbed/sympy/combinatorics/perm_groups.py line 2155
- Callee: degree (Property)
  Location: /testbed/sympy/combinatorics/perm_groups.py line 1292
- Callee: enumerate (Class)
  Location: /opt/miniconda3/envs/testbed/lib/python3.9/site-packages/pyright/dist/dist/typeshed-fallback/stdlib/builtins.pyi line 1327
- Callee: issubset (Method)
  Location: /opt/miniconda3/envs/testbed/lib/python3.9/site-packages/pyright/dist/dist/typeshed-fallback/stdlib/builtins.pyi line 1271
\end{lstlisting}
\end{LspExample}

\begin{LspExample}{get\_hover}
\begin{lstlisting}
<function=lsp_tool>
  <parameter=command>get_hover</parameter>
  <parameter=file_path>/testbed/lib/matplotlib/dates.py </parameter>
  <parameter=line>779</parameter>
</function>
\end{lstlisting}

\lspseparator 

\begin{lstlisting}
Execution output of [lsp_tool]:
[status_code]: 
 success 
[Result]: 
Hover Info:
(variable) tickdate: NDArray[Any]
\end{lstlisting}
\end{LspExample}

\begin{LspExample}{get\_document\_symbols}
\begin{lstlisting}
<function=lsp_tool>
<parameter=command>get_document_symbols</parameter>
<parameter=file_path>/testbed/sympy/sets/conditionset.py</parameter>
</function>
\end{lstlisting}

\lspseparator 

\begin{lstlisting}
[status_code]:
 success
[Result]:
Document Symbols:
- ConditionSet (Class) at (line 22 to 224)
  - __new__ (Method) at (line 91 to 165)
    - sig (Function) at (line 146 to 147)
  - free_symbols (Method) at (line 171 to 174)
  - bound_symbols (Method) at (line 176 to 179)
  - _contains (Method) at (line 181 to 198)
    - ok_sig (Function) at (line 182 to 190)
  - as_relational (Method) at (line 200 to 206)
  - _eval_subs (Method) at (line 208 to 224)
\end{lstlisting}
\end{LspExample}

\begin{LspExample}{get\_document\_highlights}
\begin{lstlisting}
<function=lsp_tool>
<parameter=command>get_document_highlights</parameter>
<parameter=file_path>/testbed/sympy/combinatorics/perm_groups.py</parameter>
<parameter=symbol>minimal_blocks</parameter>
<parameter=line>4354</parameter>
</function>
\end{lstlisting}

\lspseparator 

\begin{lstlisting}
[status_code]:
 success
[Result]:
Found 2 related highlight(s) in this document:
- (Module) at (at line 2133)
- (Module) at (at line 4354)
\end{lstlisting}
\end{LspExample}

\begin{LspExample}{get\_references}
\begin{lstlisting}
<function=lsp_tool>
<parameter=command>get_references</parameter>
<parameter=file_path>/testbed/sympy/combinatorics/perm_groups.py</parameter>
<parameter=symbol>minimal_blocks</parameter>
<parameter=line>4354</parameter>
</function>
\end{lstlisting}

\lspseparator 

\begin{lstlisting}
[status_code]:
 success
[Result]:
Found 5 reference(s) across 2 file(s):

/testbed/sympy/combinatorics/perm_groups.py:
  - (at line 2133)
  - (at line 4354)

/testbed/sympy/combinatorics/tests/test_perm_groups.py:
  - (at line 516)
  - (at line 519)
  - (at line 522)
\end{lstlisting}
\end{LspExample}

\section{LLM Prompts}\label{app:prompts}

\begin{PromptBoxNew}[p:lsp_filtering_sp]{System Prompt for LSP-Augmented Trajectory Filtering}
\begin{lstlisting}
You are an expert code analysis assistant tasked with evaluating the effectiveness of LSP (Language Server Protocol) tools in helping AI agents solve complex software engineering problems.

This is the trajectory:{trajectory}

## Your Task
Analyze the provided trajectory where an AI agent attempts to solve a coding problem from the SWE-bench dataset. Your focus is to judge whether the `lsp_tool` really brings **positive contributions** to the problem-solving process or not.
If there are any errors in the return result of lsp_tool, or if a meaningless call is made (the return result does not contain useful information), please immediately output 'no' (i.e. lsp_tool is not helpful for this trajectory)
Please think carefully and maintain an objective, rational analysis. You will be punished if you are overly biased.

## Context: The lsp_tool
The `lsp_tool` is a Pyright-based code intelligence tool that provides:
- **Semantic code understanding**: Goes beyond text search by analyzing AST and symbol tables
- **Navigation capabilities**: `get_definition`, `get_type_definition`, `get_references`
- **Code structure analysis**: `get_document_symbols`, `get_workspace_symbols`
- **Call relationship analysis**: `get_call_hierarchy`, `get_incoming_calls`, `get_outgoing_calls`
- **Contextual information**: `get_hover` for docstrings and type information

This contrasts with simpler tools like `grep`, `sed`, or `cat` which only perform text-based operations.

## Evaluation Framework

### 1. Positive Indicators (Evidence that lsp_tool is helpful)
Carefully identify instances where lsp_tool:

**a) Efficient Navigation**
- Successfully locates function/class definitions across multiple files
- Finds symbol references faster than grep-based approaches
- Discovers code relationships that would be hard to find manually

**b) Accurate Code Understanding**
- Provides structural overview via `get_document_symbols` before diving into details
- Uses `get_workspace_symbols` to quickly locate relevant code entities
- Employs `get_references` to understand how a function/class is used across the codebase

**c) Strategic Analysis**
- Uses `get_call_hierarchy` to understand function dependencies
- Leverages semantic information to make informed decisions
- Follows code flow using `get_definition` chains

**d) Problem-Solving Efficiency**
- Reduces the number of file reads needed
- Avoids blind searching through irrelevant files
- Quickly identifies the right files to modify

### 2. Negative Indicators (Evidence of ineffective usage)
Also identify cases where:

**a) Tool Misuse**
- Uses lsp_tool when simple `grep` or `cat` would suffice
- Makes redundant lsp_tool calls that don't add new information
- Uses wrong commands (e.g., `get_workspace_symbols` with overly generic queries)

**b) Failed Tool Calls**
- LSP commands fail and the agent doesn't adapt
- Symbol not found errors that waste time
- Incorrect parameters leading to errors

**c) Ignored Results**
- Agent calls lsp_tool but doesn't utilize the returned information
- Makes decisions that contradict lsp_tool findings
- Proceeds with manual file exploration after lsp_tool already provided the answer

**d) Inefficiency**
- Over-reliance on lsp_tool when the problem is straightforward
- Using multiple lsp_tool calls where one would suffice
- Mixing lsp_tool with redundant text-based searches

### 3. Comparative Analysis
Compare scenarios:
- **With lsp_tool**: How does the agent locate code? How many steps?
- **Without lsp_tool (hypothetical)**: How would the agent likely proceed using only grep/cat?
- Does lsp_tool provide a **decisive advantage** in understanding code structure?

## Output Format

Provide your analysis in the following structured format:

### Executive Summary (2-3 sentences)
Provide a clear verdict: Does lsp_tool bring positive contributions? Is it helpful, neutral, or counterproductive?

### Detailed Analysis

#### Section A: Positive Contributions
If have, for each positive use case found:
```
**Use Case #X: [Brief Title]**
- **Step Number**: [which step in the trace]
- **LSP Command**: [which command was used]
- **What it found**: [summarize the output]
- **How it helped**: [explain the benefit]
- **Efficiency gain**: [compare with alternative approaches]
```

#### Section B: Ineffective/Problematic Usage
If have, for each problematic use case:
```
**Issue #X: [Brief Title]**
- **Step Number**: [which step]
- **Problem Type**: [misuse/failed call/ignored result/inefficiency]
- **What went wrong**: [describe the issue]
- **Impact**: [how it affected problem-solving]
```

#### Section C: Key Statistics
Provide quantitative measures:
- Total lsp_tool calls: X
- Successful calls: Y (Z%)
- Calls that directly contributed to solution: N
- Failed/error calls: M
- Redundant/unnecessary calls: K

#### Section D: Critical Moments
If have, identify 1-3 pivotal moments where lsp_tool either:
- **Made a breakthrough**: Enabled the agent to find the root cause
- **Caused confusion**: Led the agent in the wrong direction
- **Was underutilized**: Could have helped but wasn't used

### Overall Assessment

**Answer**: \\boxed{{yes or no}}

**Recommendations**: 
- What could be improved in how the agent uses lsp_tool?
- Are there missed opportunities where lsp_tool should have been used?
- Should any usage patterns be avoided?

## Important Guidelines

1. **Be Evidence-Based**: Every claim must reference specific step numbers and command outputs
2. **Be Objective**: Don't assume lsp_tool is good or bad; let the evidence speak
3. **Consider Context**: A failed lsp_tool call isn't inherently bad if the agent adapts well
4. **Think Counterfactually**: Would the agent have struggled more without lsp_tool?
5. **Focus on Problem-Solving**: The ultimate question is: did it help solve the actual issue?
6. **Note Tool Synergy**: Sometimes lsp_tool + grep together are better than either alone

## Special Attention Areas

- Look for cases where `get_workspace_symbols` quickly locates the right class/function
- Check if `get_document_symbols` helps understand file structure before editing
- Observe whether `get_references` reveals unexpected usage patterns
- See if `get_definition` chains help trace code flow
- Notice when the agent **should have used** lsp_tool but didn't

Begin your analysis now.

\end{lstlisting}
\end{PromptBoxNew}

\end{document}